\documentclass[useAMS,usenatbib]{mn2e}
\usepackage{times}
\usepackage{graphicx}

\newcommand\tablenotemark[1]{% 
 \rlap{$^{\mathrm #1}$}% 
}% 

\title[Collision Products of Triple-Star Mergers]{Modelling Collision Products of Triple-Star Mergers}
\author[J. C. Lombardi, Jr., A. P. Thrall, J. S. Deneva, S. W.
Fleming, and P. E. Grabowski]{J. C. Lombardi, Jr.,\thanks{E-mail: lombardi@vassar.edu (JCL); asthrall@vassar.edu (APT); iudeneva@vassar.edu (JSD); scfleming@vassar.edu (SWF); pagrabowski@vassar.edu (PEG)}
A. P. Thrall,\footnotemark[1]
J. S. Deneva,\footnotemark[1]
S. W. Fleming,\footnotemark[1]
and P. E. Grabowski\footnotemark[1]\\
Department of Physics and Astronomy, Vassar College,
 124 Raymond Avenue, Poughkeepsie, NY 12604, USA}
\begin{document}

%\date{Accepted 1988 December 15. Received 1988 December 14; in original form 1988 October 11}

\pagerange{\pageref{firstpage}--\pageref{lastpage}} \pubyear{2003}

\maketitle

\label{firstpage}

\begin{abstract}
In dense stellar clusters, binary-single and binary-binary encounters
can ultimately lead to collisions involving two or more stars during a
resonant interaction.  A comprehensive survey of multi-star collisions
would need to explore an enormous amount of parameter space, but here
we focus on a number of representative cases involving low-mass (0.4,
0.6, and $0.8{\rm M}_\odot$) main-sequence stars.  Using both Smoothed
Particle Hydrodynamics (SPH) calculations and a much faster fluid sorting
software package ({\sevensize MMAS}), we study scenarios in which a
newly formed product from an
initial collision collides with a third parent star.   By varying the
order in which the parent stars collide, as well as the orbital
parameters of the collision trajectories, we investigate how factors
such as shock heating affect the chemical composition and structure
profiles of the collision product.  Our simulations and models
indicate that the distribution of most chemical elements within the
final product is not significantly affected by the order in which the
stars collide, the direction of approach of the third parent star, or
the periastron separations of the collisions.  Although the exact
surface abundances of beryllium and lithium in the product do depend
on the details of the dynamics, these elements are always
severely depleted due to mass loss during the collisions.  We find
that the sizes of the products, and hence their collisional cross
sections for subsequent encounters, can be sensitive to the order and
geometry of the collisions.  For the cases that we consider, the
radius of the product formed in the first (single-single star)
collision ranges anywhere from roughly 2 to 30 times the
sum of the radii of its parent stars.  The size of the final product
formed in our triple-star collisions is more difficult to determine,
but it can easily be as large or larger than a typical red giant.
Although the vast
majority of the volume in such a product contains diffuse gas that
could be readily stripped in subsequent interactions, we nevertheless
expect the collisional cross section of a newly formed product to be
greatly enhanced over that of a thermally relaxed star of the same
mass.
Our results also help
establish that the algorithms of {\sevensize MMAS} can quickly reproduce the
important features of our SPH models for these collisions, even when one of the parent
stars is itself a former product.
\end{abstract}

\begin{keywords}
stars: chemically peculiar -- globular clusters: general -- galaxies:
star clusters -- hydrodynamics -- blue stragglers -- stars: interiors
\end{keywords}

\section{Introduction\label{introduction}}

\subsection{Motivations}

One exciting aspect of dense stellar systems is the simultaneous
importance of three principal areas of stellar astrophysics:  
dynamics, evolution, and hydrodynamics.  Many
simulation codes focus on one of these areas and have often
been lifelong works in progress.  The first attempts at unifying
these treatments into a coherent model to describe clusters have
begun only recently.
In June 2002, specialists in stellar dynamics, stellar evolution,
hydrodynamics, cluster observation, visualization, and computer
science gathered at the American Museum of Natural History in New York
City to begin discussing a framework for MOdelling DEnse STellar
systems, without having to MODify Existing STellar codes extensively.
The workshop-style meeting, organized by P.\ Hut and M.\ Shara, became
known as MODEST-1 \citep{hut03}.  The second such meeting, MODEST-2,
was organized by S.\ Portegies Zwart and P.\ Hut and held in December
2002 at the Anton Pannekoek Institute in Amsterdam \citep{sil03}.
From MODEST-2, a set of eight ``working groups'' were established,
each focusing on a different aspect of the MODEST
endeavour.\footnote{See {\tt http://www.manybody.org/modest.html}.} 
Attempting to integrate stellar dynamics, evolution, and hydrodynamics codes
into one fully functional package will be challenging,
largely because each area
treats stellar properties that evolve on different time-scales.
However, by combining these areas, we will be able to better
model the origins, dynamics, evolution, and death of globular
clusters, galactic nuclei, and other dense stellar systems.

In this paper, our focus is on modelling hydrodynamic
interactions between stars.  The goal is to develop a
software module for quickly generating collision product models,
ultimately for any type of stellar collision, that could be incorporated into
simulations of dense star clusters.  \citet{lom02} presented an appropriate
formulation for treating parabolic single-single star collisions between
low-mass main-sequence stars.  Here we extend that study to include situations
in which one of the parent stars is itself a thermally {\it un}relaxed
collision product.  Such scenarios can occur during binary-single or
binary-binary interactions, when the time between collisions is much
less than the thermal relaxation time-scale.

Stellar collisions and mergers can strongly
affect the overall energy budget of a cluster and even alter the
timing of important dynamical phases such as core
collapse. Furthermore, hydrodynamic interactions are
believed to produce a number of non-canonical objects, including blue
stragglers, low-mass X-ray binaries, recycled pulsars, double neutron
star systems, cataclysmic variables, and contact binaries.  Such stars
and systems are among the most challenging to model, but they are also
among the most interesting observational markers.  Blue
stragglers, for example, exist on an extension of the main-sequence,
but beyond the turnoff point.  Blue stragglers are therefore appropriately
named, as they are more blue than the
remaining ordinary main-sequence stars, and, compared to other
stars of similar mass, are straggling behind in their evolution.
This aberration from the common path of stellar evolution is believed
to be due to mass transfer or merger in a binary system, or from the
direct collision of two or more main-sequence stars \citep[for a
review, see][]{bai95}.  Predicting the
numbers, distributions, and other observable characteristics
of stellar exotica will be
essential for detailed comparisons with observations.

\subsection{Stellar dynamics and stellar evolution}

Stellar dynamics codes determine the motions of stars.  The primary approaches to evolving clusters or galactic nuclei dynamically
are direct N-body integrations
\citep[e.g.,][]{1997A&A...328..130P,hur01}, solving the Fokker-Planck
equation \citep[e.g.,][]{2000ApJ...535..759T}, Monte Carlo approaches
\citep[e.g.,][]{wat00,joshi00,2001ApJ...550..691J,fre02,gie01,gie03}, and gaseous models
\citep[e.g.,][]{1995MNRAS.272..772S}.
For a review of the ongoing {\sevensize NBODY} effort for accurate N-body simulations, see \cite{1999PASP..111.1333A}; for a general review of cluster dynamics, see \cite{1997A&ARv...8....1M} and \cite{2003gmbp.book.....H}.

The most important quantities that a stellar evolution software module can
provide to a dynamics module are the stellar masses, as well as the
stellar radii if collisions are included.  At least in principle,
these results could come from a live
(i.e., concurrent with the cluster dynamics)
stellar evolution calculation, from fitting formulae, or from interpolation
among prior calculations.  Due to the large number of stars, it would
be wasteful to expend a considerable amount of time for a live
computation of each star's evolution.  For the ordinary
stars whose evolution is not perturbed by an event such as a
collision, it is much more efficient and entirely appropriate to
interpolate among, or to use analytic fitting formulae based upon,
previously calculated evolutionary tracks.  The parameter space
associated with non-canonical stars, however, is too enormous to
be adequately covered by interpolation or fitting formulae, and it will ultimately be
necessary to invoke a full stellar evolution calculation in parallel
with the stellar dynamics for such stars.  Although live stellar
evolution calculations have not yet been combined with stellar
dynamics codes, some parametrized codes, such as {\sevensize SeBa} \citep{zwart96},
{\sevensize SSE} \citep{hur00}
and {\sevensize BSE} \citep{hur02b}, have been successfully integrated
\citep{1997A&A...328..130P,2001MNRAS.321..199P,2002ApJ...571..830S}.

When physical collisions between stars are modelled in a cluster calculation
or a scattering experiment,
it is usually done using a method known as ``sticky
particles,'' in which a collision product is given a mass equal to the
combined mass of the two parent stars and a velocity determined by
momentum conservation.  In a collision between two main-sequence
stars, for example, the result would be modelled as a rejuvenated,
thermally relaxed main-sequence star.  This simple method
is a reasonable first approximation for many situations.
However, there are important characteristics of collision products
that are neglected, including their rapid rotation, peculiar
composition profiles, and enhanced sizes due to shock heating.  If the
thermal relaxation time-scale of the collision product is much less
than the time until its next collision, then it is appropriate to
assume the product becomes instantaneously thermally
relaxed, as is done in the classic simulations of \cite{qui90}.  This approximation becomes questionable when
the collision has been mediated by a binary, as there is then at least
one star in the immediate vicinity of the collision product and the
likelihood of a subsequent collision will depend sensitively on the
product's thermally {\it un}relaxed size.

Binaries are subject to enhanced collision rates for two primary
reasons: (1) their collision cross section depends on the semi-major
axis of the orbit, as opposed to the radius of a single star, and (2)
due to mass segregation, binaries tend to be found in the core of the
cluster, the densest and most active region.  In clusters with a binary
fraction exceeding about 20 per cent, binary-binary collisions are
expected to occur more frequently than single-single and binary-single
collisions combined \citep{1996MNRAS.281..830B}.
It is probably not uncommon for binary fractions to be this large:
the inner core of NGC 6752, for example, is thought to have
a binary fraction in the range 15--38 per cent \citep{rub97}.

Binary populations can lead to complex and chaotic resonant
interactions.  These interactions tend to exchange energy between the
binaries and the other stars in the cluster, and therefore are
critical in determining its dynamics and observable
characteristics \citep{hut92,1994ApJ...431..231V}.  A star intruding on
a binary could, depending on parameters such as 
the separation of the binary and the velocity of the incoming star, 
escape to infinity, destroy the binary,
form a new binary with a star from the original, or form a triple.
The outcomes of three body
encounters can be categorized using a nomenclature based upon
typical atomic processes \citep[see][who introduced the use of
terms like ``ionization'' and ``exchange,''
to describe resultant scenarios]{heg75}.  See \citet{hur02} for an
informative narration of the intimate
interactions, usually involving binaries, that stars regularly undergo
during cluster evolution.

The {\sevensize STARLAB} computing environment is a very useful tool for
modelling and analyzing all types of stellar phenomena.  The general technique for using {\sevensize STARLAB} to determine
the cross sections or branching ratios for the various outcomes
of binary interactions is presented by \citet{mcm96}.
They also highlight an example set of cases in
which a $0.6 {\rm M}_\odot$ star intrudes upon a binary with a
$0.8 {\rm M}_\odot$
primary and a $0.4 {\rm M}_\odot$ secondary.  Their assumed mass radius
relation, $M/{\rm M}_\odot =R/{\rm R}_\odot$, is appropriate for thermally relaxed
main sequence stars and is applied both to the parent stars and any
collision products.  They find that for binaries with semimajor axes
of 0.2, 0.1, 0.05, and 0.02 AU, triple-star mergers comprise about 1, 2,
5, and 15 per cent, respectively, of all merger events.  As the results of
the present paper will help show, we expect that
accounting for the enhanced, thermally unrelaxed size of the first collision
product will greatly increase these percentages as well as the range of
semimajor axes in which triple-star mergers are significant.

Simulations of moderately dense galactic nuclei initially
containing solar-mass main-sequence stars demonstrate that
runaway mergers can readily produce stars with masses $\ga 100 {\rm M}_\odot$.
These massive stars then undergo further mergers to produce seed black holes
 with masses
as large as $\sim10^3{\rm M}_\odot$ \citep{qui90}.  This process may be
responsible for massive black holes at the centres of most galaxies,
including our own.  For star clusters, recent N-body simulations
reveal that runaway mergers can
lead to the creation of central black holes within a few million years
\citep[e.g.,][]{por99,2002ApJ...576..899P}.  With the help of Monte
Carlo
simulations, \citet{ras03} show that the runaway process will
occur in a typical cluster with a relaxation timescale
less than about 30 Myr.  
Observational evidence for a possible intermediate-mass black hole in M15 has been recently reported
by \cite{ger02}, although the data is more reasonably modelled with a large
concentration of stellar-mass compact objects \citep{baum03,ger03}.

\subsection{Stellar hydrodynamics \label{sh}}

Mass loss and expansion due to shock heating when two stars collide
are examples of hydrodynamical processes that can ultimately affect
the future evolution of the cluster.   Mostly using the Smoothed
Particle Hydrodynamics (SPH, see \S\ref{sph}) method, numerous
scenarios of stellar collisions and mergers have been simulated in
recent years, including collisions between two main-sequence stars
\citep{ben87,lai93,lom96,oue98,san97,sil97,1997ApJ...487..290S,sil01,sil02,fre03},
collisions between a giant star and compact object \citep{ras91}, and
common envelope systems \citep{ras96,ter94,ter95,san98,san00}.
The first published SPH calculations of three-body encounters
were done by \citet{cle90}, who performed over 100 very low resolution
simulations and implemented a mass-radius relation appropriate for
white dwarfs.  Other three- and four-body interaction simulations
include binary-binary encounters among $n=1.5$
polytropes \citep{goo91} as well as neutron star -- main-sequence binary
encounters with a neutron star, main-sequence or white dwarf intruder
\citep{dav94}.  See \citet{ras99} for more information concerning the
use of SPH in stellar collisions, and see \citet{sha02} for a
qualitative overview of the progress in stellar collision research.

If the structure and composition profiles of colliding stars were
available (perhaps from a live stellar evolution calculation) during
a cluster simulation, then the sticky particle method could be
replaced by a more detailed hydrodynamics module.  SPH calculations
could then, at least in principle, be run on demand within this
cluster simulation in order to determine the orbital trajectory of
the product(s), as well as their structure and chemical composition
distributions.  However, at least $10^5$ SPH fluid particles may be
necessary to allow an accurate treatment of the subsequent evolution
of collision products \citep{sil02}.  The trouble, therefore, is that
the integration of just a single interaction could consume hours, days
or even weeks of computing time (depending on the initial conditions,
desired resolution, and available computational resources).  Although
the use of equal-mass particles, or the more accurate SPH equations of
motion derived by \citet{2002MNRAS.333..649S}, or both, could decrease
the total number of particles required, it is still currently impractical to
implement a full hydrodynamics calculation for every close stellar
encounter during a cluster simulation.

One approach for incorporating strong hydrodynamic interactions and
mergers into a grand simulation of a cluster, already successfully
implemented by \citet{fre02} in the context of galactic nuclei, is to
interpolate between the results of a set of previously completed SPH
simulations. The SPH database of Freitag \& Benz treats all types of
hyperbolic collisions between main-sequence stars (mergers, fly-bys
and cases of complete destruction), while also varying the parent star
masses as well as the eccentricity and periastron separation of their
initial orbit.  The tremendous amount of parameter space surveyed
precludes having high enough resolution to determine the detailed
structure and composition profiles of the collision products for all
cases; however, critical quantities such as mass loss and final
orbital elements can be determined accurately.

A second possibility is to forgo hydrodynamics simulations and instead
model collision products by physically motivated algorithms and
fitting formulae that sort the fluid from the parent stars
\citep{lom02}.  One advantage of such an approach is that it can
handle cases in which one or both of the parent stars is itself a
former collision product (with chemical and structural profiles that
are substantially different than that of a standard isolated star of
similar mass and type).

In this paper, we use both SPH calculations and a much faster fluid
sorting algorithm to study scenarios in which a newly formed collision
product collides with a third parent star.  
By varying the order and orbital parameters of the collision,
we investigate how factors such as shock heating affect
the chemical composition and structure profiles of the collision
product.  Section \ref{procedure} presents our procedures and
numerical methods, both for our SPH calculations (\S\ref{sph}) and our
fluid sorting algorithm (\S\ref{sorting}).  SPH
results are presented in \S\ref{sphresults}, and then compared to the
results of our fluid sorting algorithm in \S\ref{mmasresults}.  In
\S\ref{discussion} we discuss our findings and possible directions for future
work.

\section{Procedure} \label{procedure}
\subsection{Smoothed Particle Hydrodynamics \label{sph}}

One means by which we generate collision product models is with the
parallel SPH code used in \citet{sil01}.  The original serial version
of this code was developed by \cite{R91}, specifically for the study
of stellar interactions such as collisions and binary mergers
\citep[see, e.g.,][]{ras91,RS92,RS94}.  Introduced by \citet{luc77}
and \citet{gin77}, SPH is a hydrodynamics method that uses a smoothing
kernel to calculate local weighted averages of thermodynamic
quantities directly from Lagrangian fluid particle positions
\citep[for a review, see][]{mon92}.  Each SPH particle can be thought
of as a parcel of gas that traces the flow of the fluid, with the
kernel providing each particle's spatial extent and the means by which
it interacts with neighbouring particles.

The SPH code solves
the equations of motion of a large number of
particles moving under the influence of both hydrodynamic and
self-gravitational forces.  All of the scenarios we investigate using SPH
involve a $0.8 {\rm M}_\odot$ parent star, represented with
12800 equal-mass SPH particles, and two $0.6 {\rm M}_\odot$ parent stars, each
represented with 9600 equal-mass particles.  Each of our SPH particles therefore
has a mass of $6.25\times10^{-5} {\rm M}_\odot$.  For comparison,
this particle mass is between the masses of the central particles
used in the $N=3\times 10^5$ and $N=10^6$ calculations of
\citet{sil02}, who used unequal-mass particles to study in detail the
outer layers of the fluid in a collision between two $0.6{\rm M}_\odot$\
main-sequence stars.  For our purposes, the use of
equal-mass particles is more appropriate, as it allows for higher
resolution in the stellar cores and does not waste computational
resources on the ejecta.

Local densities and hydrodynamic forces at each particle position are
calculated by appropriate summations over $N_N$ nearest neighbours.
The size of each particle's smoothing kernel determines the local
numerical resolution and is adjusted during each time step to keep
$N_N$ close to a predetermined value, 48 for the present calculations.
Neighbour lists for each particle are recomputed at every iteration
using a linked-list, grid-based parallel
algorithm \citep{1988csup.book.....H}.

The hydrodynamic forces acting on each particle include an artificial
viscosity contribution that accounts for shocks.  As in \citet{sil01},
we adopt the artificial viscosity form proposed by \cite{bal95}, with
$\alpha=\beta=5/6$, $\eta^2=0.01$, and $\eta^{\prime^2}=10^{-5}$.
This form treats shocks well and has the tremendous advantage that it
introduces only relatively small amounts of spurious shock heating and
numerical viscosity in shear layers \citep{lom99}.

A number of physical quantities are associated with each SPH particle,
including its mass, position, velocity, and entropic variable $A$.
Here we adopt a monatomic ideal gas equation of state,
appropriate for the stars in our mass range.  That is, $P=A\rho^{\gamma}$,
where the adiabatic index
$\gamma=5/3$ with $P$ and $\rho$ being the pressure and density,
respectively.  The entropic variable is closely related (but not
equal) to specific entropy: both of these quantities are conserved in
the absence, and increase in the presence, of shocks.

Our code uses an FFT-based convolution method to calculate
self-gravity.  The fluid density is placed on a zero-padded, 3D grid
by a cloud-in-cell method, and then convolved with a kernel function
to obtain the gravitational potential at each point on the grid.
Gravitational forces are calculated from the potential by finite
differencing, and then interpolated for each particle using the same
cloud-in-cell assignment scheme.  For each collision simulation
in this paper,
the number of grid cells is $256^3$.  The ejecta leaving the grid
interact with the enclosed mass simply as if it were a monopole.

Following the same approach as in \citet{sil01}, we begin by using a
stellar model from the Yale Rotational Evolution Code ({\sevensize
YREC}) to help generate SPH models of the parent stars.
We focus on collisions involving 0.8 and
$0.6 {\rm M}_\odot$ main-sequence stars, with a primordial helium abundance
$Y=0.25$ and metallicity $Z=0.001$.  Using {\sevensize YREC}, these
stars were evolved with no rotation to an age of 15 Gyr, the amount of
time needed for the $0.8{\rm M}_\odot$ star to reach turnoff.  The total
helium mass fractions for the $0.6$ and $0.8 {\rm M}_\odot$ parent
stars are 0.286 and 0.395, and their radii are $0.517$ and
$0.955 {\rm R}_\odot$, respectively.  See figs.\ 1 and 2 of \citet{lom02} for thermodynamic
and composition profiles of the parent stars presented as a function
of enclosed mass, as determined by YREC.

To generate our SPH models, we use a Monte Carlo
approach to distribute particles according to the
desired density distribution, determining values of $A$ for each SPH
particle from its position.  To minimize numerical noise, an
artificial drag force is implemented, with artificial viscosity turned off,
to relax each SPH parent model
to the equilibrium configuration used to initiate the collision
calculations.  Fourteen different chemical abundance profiles are
available from the {\sevensize YREC} parent models to set the
composition of the SPH particles.  The abundances of an SPH particle
are assigned according to the amount of mass enclosed by an isodensity
surface passing through that particle in the relaxed configuration.

\begin{figure}
\includegraphics[width=84mm]{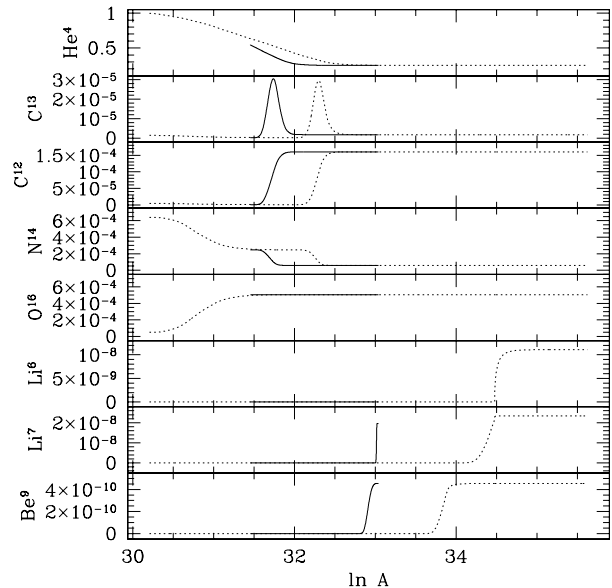}
% refpar2:
\caption{Fractional chemical abundances (by mass) of the elements He$^4$,
C$^{13}$, C$^{12}$, N$^{14}$, O$^{16}$, Li$^6$, Li$^7$, and Be$^9$ versus $\ln A$, where the entropic variable $A\equiv P/\rho^{5/3}$ is in cgs units, for
our $0.6 {\rm M}_\odot$ (solid curve) and $0.8 {\rm M}_\odot$ (dotted curve) parent stars, as determined by {\sevensize YREC}.
}
\label{refpar2}
\end{figure}

Fig.~\ref{refpar2} plots fractional chemical abundances
(by mass) versus $\ln A$ in each parent star in its relaxed SPH
configuration.  Note that the dense core of the turnoff star is
at a smaller $A$, and its diffuse outer layers are at a larger $A$,
than all of the fluid in the $0.6 {\rm M}_\odot$ star, which has direct
consequences for the hydrodynamics of collisions involving these
stars.  Also note that lithium and beryllium exist only in the
outermost layers of the parent stars.

%%%%%%%%%%%%%%%%%%%%%%%%%%%%%%%%%%%%%%%%%%%%%%%%%%%%%%%%%%%%%%%%%%%%%%%%%
% FIGURE 1
%%%%%%%%%%%%%%%%%%%%%%%%%%%%%%%%%%%%%%%%%%%%%%%%%%%%%%%%%%%%%%%%%%%%%%%%%

We focus on triple-star collisions, modelling each collision
separately and in succession.  We do not consider fly-bys or grazing
collisions in our SPH calculations: all of our collisions lead to mergers.  We neglect
any direct or indirect effects, including tidal forces, that the third
star may have on the dynamics of the first collision.  We assume
that the second collision occurs before the first collision product
thermally relaxes: a reasonable approximation since contraction to the
main-sequence occurs on a thermal time-scale, lasting at least $\sim
10^6$ yr for non-rotating products and as long as $\sim 10^8$ yr for
rapidly rotating products \citep{1997ApJ...487..290S,sil01}, much
longer than the typical time between collisions in some binary-single or
binary-binary interactions (but see \S\ref{future}).

The orbital trajectory in all our collisions is taken to be parabolic.
This is clearly not appropriate for galactic nuclei, where collisions
are typically hyperbolic.  However, in globular clusters, the velocity
dispersion is only $\sim 10$ km s$^{-1}$, much less than the 600 km
s$^{-1}$ escape speed from the surface of our $0.8 {\rm M}_\odot$
turnoff star, and
hence all single-single star collisions are essentially parabolic.
For collisions involving binaries (even including some hard binaries), the
escape speed can still be large compared to the effective relative velocity at
infinity.  For example, consider two $0.8 {\rm M}_\odot$ turnoff stars in a
circular orbit of radius 0.05 AU in a globular cluster.
This is a hard binary,
as each star moves with a velocity of about 60
km s$^{-1}$ with respect to the center
of mass of the binary, a speed significantly larger than the cluster
velocity dispersion.
Yet, the effective relative velocity at infinity for
a collision between one of the binary components
and an intruder would typically not be much more than the orbital
speed, and therefore still significantly less than 
the escape speed from our turnoff star.  We
therefore expect collisions between a slow intruder and a binary
component to be close to parabolic not only for all soft binaries, but
also for some (moderately) hard binaries

For the first single-single star collision, the stars are initially
non-rotating and separated by 5 $R_{TO}$, where $R_{TO}=0.955{\rm R}_\odot$
is the radius of our turnoff star.  The initial velocities are
calculated by approximating the stars as point masses on an orbit with
zero orbital energy and a periastron separation $r_p$.  A Cartesian
coordinate system is chosen such that these hypothetical point masses
of mass $M_1$ and $M_2$ would reach periastron at positions
$x_i=(-1)^{i}(1-M_i/(M_1+M_2))r_p$, $y_i=z_i=0$, where $i=1,2$ and
$i=1$ refers to the more massive star.  The orbital plane is chosen to
be $z=0$.  With these choices, the centre of mass resides at the
origin.  For the first collisions, the gravity grid maintains a fixed
spatial extent from -4$R_{TO}$ to +4$R_{TO}$ along each dimension.

%%%%%%%%%%%%%%%%%%%%%%%%%%%%%%%%%%%%%%%%%%%%%%%%%%%%%%%%%%%%%
% FIGURE 2
\begin{figure}
\includegraphics[width=84mm]{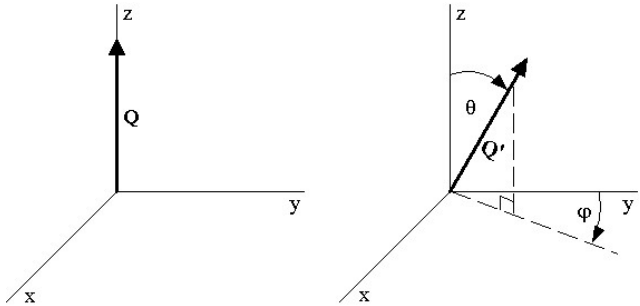}
\caption{Before initiating the second collision, we rotate the final position
  and velocity vectors of each SPH particle from
the first collision: first by an angle $\theta$ around the $x$-axis
  and then by an angle $\phi$ around the $z$-axis.  The above figure shows how a vector ${\bf Q}$
initially pointing along the $z$-axis is transformed into the new
vector ${\bf Q^\prime}$ by these rotations.
}
\label{rotate}
\end{figure}
%%%%%%%%%%%%%%%%%%%%%%%%%%%%%%%%%%%%%%%%%%%%%%%%%%%%%%%%%%%%%

For the second collision, we want to control the relative orientation
of the first collision product's rotation axis and the orbital plane
(or, equivalently, the direction of approach of the third parent star).
To do so, we begin with the final state of the first collision and
make two rotations to its particle positions and velocities, through the angles
$\theta$ and $\phi$.
More specifically, the first rotation is
clockwise through an angle $\theta$ about the $x$ axis, while
the second rotation is clockwise through an angle $\phi$ about the $z$
axis (see Fig.~\ref{rotate}).
Finally, the particle positions and velocities are uniformly shifted
parallel to the $x$-$y$ plane, and the third star
is introduced such that the system's centre of mass will
remain at the origin and the periastron positions (in the two body
point mass approximation) will occur on the $x$ axis. 
In order to allow the bulk of the fluid to remain within
the gravity grid, the grid is extended up to a full width of
$10R_{TO}$ in the $x$ and $y$ directions for some of the second collisions.

We use the same iterative procedure as \citet{lom96} to determine the
bound and unbound mass.  SPH structure and composition profiles presented
in this paper result from averaging in 100 equally sized bins in the
bound mass.  Unfortunately, it is extremely difficult to use SPH
simulations to specify the equilibrium structure of the outermost few
per cent of mass in any collision product.  Some SPH particles,
although gravitationally bound, are ejected so far from the system's
centre of mass that it would take many dynamical time-scales for them
to rain back onto the central product and settle into equilibrium.
Our requirement for stopping an SPH calculation is that the entropic
variable $A$, when averaged over isodensity surfaces, increases
outward over at least the inner 95 per cent of the bound mass in the
first collision product, and at least 92 per cent in the second
collision product.  Many calculations are run longer in order to
confirm that no rapid changes are still occurring in the structure and
chemical composition distributions.

\subsection{Make Me A Star\label{sorting}}

The results of parabolic collisions between low-mass main-sequence
stars can be well explained by simple physical arguments.  To a good
approximation, the fluid from the parent stars sorts itself such that
fluid with the lowest values of $A$ sinks to the core of the collision
product while the larger $A$ fluid forms its outer layers. Therefore,
the interior structure and the chemical composition profiles of the
collision product can be predicted accurately using simple algorithms,
instead of hydrodynamic simulations.  Based on these ideas,
\citet{lom02} have recently created a publicly available software
package, dubbed Make Me A Star ({\sevensize MMAS}).\footnote{See {\tt
http://faculty.vassar.edu/lombardi/mmas/}.}  This package
produces collision product models close to those of an SPH code in
considerably less time, while still accounting for shock heating, mass
loss, and fluid mixing.

Sorting the shocked fluid according to its entropic variable $A$ gives
the $A$ profile of the collision product as a function of the mass $m$
enclosed inside an isodensity surface.  In the case of the
non-rotating products formed in head-on ($r_p=0$) collisions, knowledge of the
$A(m)$ profile is sufficient to determine the pressure $P(m)$, density
$\rho(m)$, and radius $r(m)$ profiles.   Using the $A$ profile
determined by sorting, {\sevensize MMAS} numerically integrates the
equation of hydrostatic equilibrium with $dm=4\pi r^2 \rho dr$ to
determine the $\rho$ and $P$ profiles, which are related through
$\rho=(P/A)^{3/5}$. The outer boundary condition is
that $P=0$ when $m=M_{MMAS}$, where $M_{MMAS}$ is the desired
(gravitationally bound) mass of the collision product.
The virial theorem provides a check of the resulting profiles.
This approach
allows for the quick generation of collision product models, without
hydrodynamic simulations, and has already been tested with single-single star
collisions.

Presented in this paper are the results from {\sevensize MMAS} for
triple-star collisions (see \S\ref{mmasresults}).  Our procedure
is simple.  We
call the {\sevensize MMAS} routine twice, using the output model from
the first collision as one of the input parent models in the second.
These {\sevensize MMAS} calculations therefore account for the
differences in shock heating that arise from changing the order, or
the periastron separations, or both, of the collisions.  In addition
to investigating all of the scenarios considered with the SPH code, we
also use {\sevensize MMAS} to examine more completely how the sizes of
products vary with the periastron separations of the collisions.
Furthermore, we include a $0.4 {\rm M}_\odot$ parent star of radius
$0.357 {\rm R}_\odot$, whose structure is determined by {\sevensize
YREC} under the same conditions described in \S\ref{sph}.

For an off-axis collision, knowledge of the specific angular momentum
distribution in the collision product is necessary
to determine its structure fully, which by itself is a challenging
problem
\citep{1968ApJ...151.1075O,1978ApJ...222..967C,1979ApJ...230..230C,1991A&A...248..435E,1995MNRAS.277.1411U}.
Although {\sevensize MMAS} outputs an approximate specific angular
momentum profile of the first collision product, we use only its
entropic variable $A(m)$ profile to help initiate the second
collision.  That is, for one of the parent stars in the second
collision, we always give as input to {\sevensize MMAS} the structure
of a non-rotating star with the desired $A(m)$ profile, a
simplification that both eases and quickens computations.
The validity of this approximation is supported by the SPH
calculations presented in \S\ref{spin}.

\citet{lom02} implemented version 1.2 of
{\sevensize MMAS}, while the results of this paper use version 1.6.
Besides cosmetic changes, the primary enhancement is that the
structure of the collision product is integrated with
a Fehlberg fourth-fifth order Runge-Kutta method.  In addition, we
have fine-tuned the fitted parameter $c_3$ from its previous value of
-1.1 to the new value -1.0, which has the effect of distributing shocks
slightly more uniformly throughout the fluid.

%%%%%%%%%%%%%%%%%%%%%%%%%%%%%%%%%%%%%%%%%%%%%%%%%%%%%%%%%%%%%%%%%%%
% SECTION 3
\section{Results\label{results}}

Table \ref{singlesingle} summarizes five single-single star
%%%%%%%%%%%%%%%%%%%%%%%%%%%%%%%%%%%%%%%%%%%%%%%%%%%%%%%%%%%%%%%%%%%
% TABLE 1
\begin{table*}
\begin{minipage}{110mm}
\caption{Summary data of six single-single star collisions.\label{singlesingle}}
\begin{tabular}{@{}ccccccccc}
\hline
Case & $M_1$ & $M_2$ & $r_p$ & $T_f$ & $R_{0.9,SPH}$\footnote{
The SPH and {\sevensize MMAS} radii 
should not be directly compared.  
The {\sevensize MMAS} 90 per cent radii are for a spherical,
non-rotating star with the same $A(m)$ profile as the collision product,
while the SPH results
account for the rotation of the product.
Furthermore, the SPH radii are measured at the time $T_f$ that the
simulation was
terminated, before all the fluid has fallen back to the product, while
the {\sevensize MMAS}
radii account for the shock heating this fluid will produce.}
 & $R^\prime_{0.9,MMAS}$\tablenotemark{\thempfootnote} & $M_{SPH}$ & $M_{MMAS}$
\\
&   [${\rm M}_\odot$] & [${\rm M}_\odot$] & [${\rm R}_\odot$] & [hr] & [${\rm R}_\odot$] & [${\rm R}_\odot$] &
[${\rm M}_\odot$] & [${\rm M}_\odot$]
\\
\hline
d  & 0.8 & 0.6 & 0.000 & 6.24  & 0.87 & 0.88 & 1.304 & 1.301\\
e  & 0.8 & 0.6 & 0.368 & 8.09  & 1.20 & 1.24 & 1.369 & 1.362\\
j  & 0.6 & 0.6 & 0.000 & 6.24  & 0.91 & 0.72 & 1.132 & 1.121\\
jk & 0.6 & 0.6 & 0.123 & 20.9  & 1.11 & 0.79 & 1.156 & 1.149\\
k  & 0.6 & 0.6 & 0.247 & 27.7  & 1.52 & 0.83 & 1.169 & 1.162\\
\hline
\end{tabular}
\end{minipage}
\end{table*}
%%%%%%%%%%%%%%%%%%%%%%%%%%%%%%%%%%%%%%%%%%%%%%%%%%%%%%%%%%%%%%%%%%%
simulations of parabolic collisions.  The table lists: the case name;
the masses $M_1$ and $M_2$ of the parent stars $i=1$ and $2$, respectively; the periastron separation
$r_p$ of the initial orbit; the stellar time $T_f$ when the
calculation was terminated; the average radius of the isodensity surface
enclosing 90 per cent of the bound mass, as determined by SPH, $R_{0.9,SPH}$,
and by {\sevensize MMAS}, $R^\prime_{0.9,MMAS}$;
and the final mass of the collision
product as calculated both by SPH, $M_{SPH}$, and by {\sevensize
MMAS}, $M_{MMAS}$.  Previous, higher resolution SPH simulations of
cases e and k \citep[see][]{sil01,lom02}
yield no significant differences from the present calculations.
For the cases in Table \ref{singlesingle}, the mass loss percentage ranges
from about 2 to 7 per cent.  Comparing the last two columns of this
table, we see that the mass loss prescription of
{\sevensize MMAS} reproduces the SPH results for the final product
mass to within 1 per cent in all six cases.

Table \ref{second} summarizes scenarios in which a collision product
%%%%%%%%%%%%%%%%%%%%%%%%%%%%%%%%%%%%%%%%%%%%%%%%%%%%%%%%%%%%%%%%%%%
%TABLE 2
\begin{table*}
%\centering
\begin{minipage}{150mm}
\caption{Summary data of collisions
between a third star and a product of a first collision.\label{second}}
\begin{tabular}{@{}cccccccccccc}
\hline
 Case & First & $M_3$ &
 $r_{p,2}$ & $\theta$ & $\phi$ & $T_f$& $T/|W|$ & $R_{0.9,SPH}$
\footnote{The same caution must be exercised when comparing the SPH
  and {\sevensize MMAS} radii as with the data of Table \ref{singlesingle}.}
& $R^\prime_{0.9,MMAS}$\tablenotemark{\thempfootnote} &
 $M_{SPH}$ & $M_{MMAS}$ 
\\
     & Product
& [${\rm
 M}_\odot$] & [${\rm R}_\odot$] & [$^\circ$] & [$^\circ$] & [hr] &
& [${\rm R}_\odot$] & [${\rm R}_\odot$] 
& [${\rm M}_\odot$] & [${\rm M}_\odot$] 
\\
\hline
%    M_1  M_2   r_p   theta psi  T_f   T/|W|   R_SPH R_MMAS M_SPH  M_MMAS 
1  &  d & 0.6 & 0.00  & 0 & 0  & 8.78 & 0.001 & 1.2 & 1.66 &1.765 & 1.747\\
2  &  j & 0.8 & 0.00  & 0 & 0  & 15.7 & 0.001 & 1.3 & 1.36 &1.769 & 1.760\\
3  &  e & 0.6 & 0.00  & 0 & 0  & 12.0 & 0.011 & 1.7 & 2.31 &1.799 & 1.778\\
4  &  e & 0.6 & 0.0955& 0 & 0  & 12.7 & 0.059 & 2.0 & 3.09 &1.851 & 1.825\\
5  &  k & 0.8 & 0.198 & 0 & 0  & 23.1 & 0.060 & 2.5 & 2.05 &1.868 & 1.862\\
6  &  k & 0.8 & 0.198 & 45& 0  & 23.1 & 0.056 & 2.4 & 2.05 &1.862 & 1.862\\
7  &  k & 0.8 & 0.00  & 0 & 0  & 21.0 & 0.005 & 1.8 & 1.48 &1.794 & 1.789\\
8  &  k & 0.8 & 0.00  & 90& 90 & 15.7 & 0.005 & 1.7 & 1.48 &1.788 & 1.789\\
9  &  k & 0.8 & 0.00  & 45& 90 & 11.5 & 0.006 & 1.7 & 1.48 &1.793 & 1.789\\
10 &  k & 0.8 & 0.00  & 45& 0  & 15.7 & 0.006 & 1.8 & 1.48 &1.794 & 1.789\\
11 &  jk& 0.8 & 0.00  & 0 & 0  & 14.8 & 0.004 & 1.5 & 1.44 &1.792 & 1.781\\
12 &  j & 0.8 & 0.505 & 0 & 0  & 23.1 & 0.077 & 2.4 & 2.25 &1.883 & 1.866\\
13 &  jk& 0.8 & 0.505 & 0 & 0  & 23.1 & 0.091 & 2.9 & 2.48 &1.893 & 1.890\\
14 &  k & 0.8 & 0.505 & 0 & 0  & 23.1 & 0.092 & 3.3 & 2.60 &1.893 & 1.902\\
15 &  k & 0.8 & 0.505 & 90& 90 & 23.1 & 0.081 & 3.2 & 2.60 &1.893 & 1.902\\
16 &  k & 0.8 & 0.505 & 45& 90 & 23.1 & 0.088 & 3.3 & 2.60 &1.893 & 1.902\\
17 &  k & 0.8 & 0.505 & 45& 0  & 23.1 & 0.090 & 3.3 & 2.60 &1.895 & 1.902\\
18 &  k & 0.8 & 0.505 & 90& 0  & 23.1 & 0.079 & 3.1 & 2.60 &1.894 & 1.902\\
19 &  k & 0.8 & 0.505 &180& 0  & 23.1 & 0.064 & 2.7 & 2.60 &1.885 & 1.902\\
20 &  k & 0.8 & 0.758 & 45& 0  & 39.3 & 0.091 & 4.2 & 2.90 &1.912 & 1.917\\
\hline
\end{tabular}
\end{minipage}
\end{table*}
%%%%%%%%%%%%%%%%%%%%%%%%%%%%%%%%%%%%%%%%%%%%%%%%%%%%%%%%%%%%%%%%%%%
from Table \ref{singlesingle} (referred to as the first collision
product) is collided with a third parent star.  The table shows: the
case number; the name of the single-single collision that yielded the
first collision product; the mass $M_3$ of the third ($i=3$) parent star; the
periastron separation $r_{p,2}$ of the second collision; the 
rotation angles $\theta$ and $\phi$ (see Fig.\ \ref{rotate}); the time $T_f$
when the calculation was terminated; the ratio of kinetic to
gravitational binding energy $T/|W|$ in the centre-of-mass frame of
the final SPH collision product model; the average radius of the isodensity
surface enclosing 90 per cent of the bound mass, as calculated by SPH,
$R_{0.9,SPH}$, and by {\sevensize MMAS}, $R_{0.9,MMAS}$;
and the mass of the product as
calculated both by SPH, $M_{SPH}$, and by {\sevensize MMAS},
$M_{MMAS}$.

All twenty cases presented in Table \ref{second} involve two $0.6 {\rm
M}_\odot$ stars and a single $0.8 {\rm M}_\odot$ star.  If mass loss
were neglected completely, the mass of the final collision product
would therefore simply be $2.0 {\rm M}_\odot$.  The SPH calculated
masses range from about 1.76 to $1.91 {\rm M}_\odot$, with the largest
mass loss occurring for cases with successive head-on collisions.

The cases in Table \ref{second} group naturally together in a variety
of ways.  Cases 1 and 2 each involve two head-on collisions.  Cases 5
and 6 differ only in the orientation of the first collision product's
spin axis, and an identical statement can be made for cases 7 through
10, as well as for cases 14 through 19.  Cases 2, 11, and 7 differ only
in the periastron separation of the first collision, as do cases 12,
13 and 14.  Also, many of the cases differ only in the periastron
separation for the second collision: for example, cases 7, 5, and 14,
as well as cases 10, 6, 17, and 20.

Even without running an SPH or {\sevensize MMAS} calculation, one can
generate a ``zeroth order'' collision product model, valid for all
twenty cases, simply by sorting the fluid of the three parent stars by
their $A$ values, with $A$ increasing from the core to the surface.
In those regions in which more than one parent star contributes,
chemical abundances can be determined by an appropriate
weighted average: the fraction of fluid with entropic variable in some
range $(A,A+\Delta A)$ that originated from any one parent star is
just equal to the fluid mass in that same range from that star divided
by the total fluid mass in that range from all three stars.

\begin{figure}
\includegraphics[width=84mm]{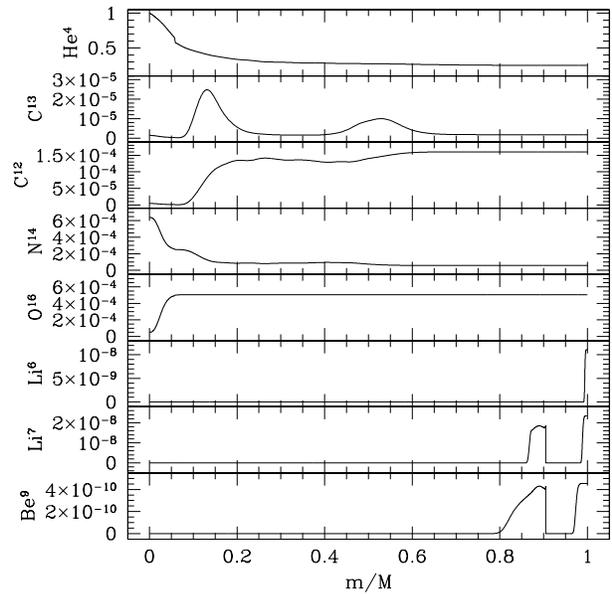}
\caption{Fractional chemical abundances (by mass) versus enclosed mass fraction
$m/M$ for a ``zeroth order'' collision product model generated by
sorting the fluid from one $0.8$ and two $0.6 {\rm M}_\odot$ stars
according to their $A$ values, without accounting for mass loss, shock heating or fluid mixing.  Here $m$ is
the mass enclosed within a surface of constant density and $M$ is the
total mass of the collision product, $2.0{\rm M}_\odot$ in this model.
}
\label{zeroth_order}
\end{figure}

Fig.~\ref{zeroth_order} shows the composition profiles resulting from
%%%%%%%%%%%%%%%%%%%%%%%%%%%%%%%%%%%%%%%%%%%%%%%%%%%%%%%%%%%%%
% FIGURE 3
%%%%%%%%%%%%%%%%%%%%%%%%%%%%%%%%%%%%%%%%%%%%%%%%%%%%%%%%%%%%%
this exercise for the merger of two $0.6 {\rm M}_\odot$ stars and a $0.8
{\rm M}_\odot$ star, with shock heating, mass loss, and fluid mixing all
completely neglected.  The innermost 6 per cent and outermost 9 per
cent of this collision product model consist of fluid that originated
entirely in the $0.8 {\rm M}_\odot$ parent star, and the profiles there
consequently mimic those of the innermost and outermost regions,
respectively, of that parent.  The profiles in the intermediate
region, where all three parent profiles contribute, can be understood
by looking back at Fig.~\ref{refpar2} and remembering that the smaller
$A$ fluid is placed deeper in the product.  For example, the C$^{13}$
near $m/M=0.13$ originated in the $0.6 {\rm M}_\odot$ star, while the higher
$A$, C$^{13}$-rich fluid peaked near $m/M=0.5$ originated mostly in
the $0.8 {\rm M}_\odot$ star.  Although some of the composition profiles
turn out to agree reasonably with our more precise SPH and {\sevensize
MMAS} calculations, the structure of the star produced by this simple
merging procedure does not.  In particular, because shock heating is
neglected, energy is not conserved during the merger, and the
resulting radius for the product is a considerable
underestimate.  The primary usefulness of the model presented in
Fig.~\ref{zeroth_order} is that it serves as a reference to help us
evaluate in what ways the shock heating,
mass loss, and fluid mixing
treated by our SPH and {\sevensize MMAS} calculations affect the
composition and structure profiles of the collision product.

% Section 3.1
\subsection{Smoothed Particle Hydrodynamics\label{sphresults}}

% Section 3.1.1
\subsubsection{Varying the collision order\label{order}}

\begin{figure}
\centering
\includegraphics[width=84mm]{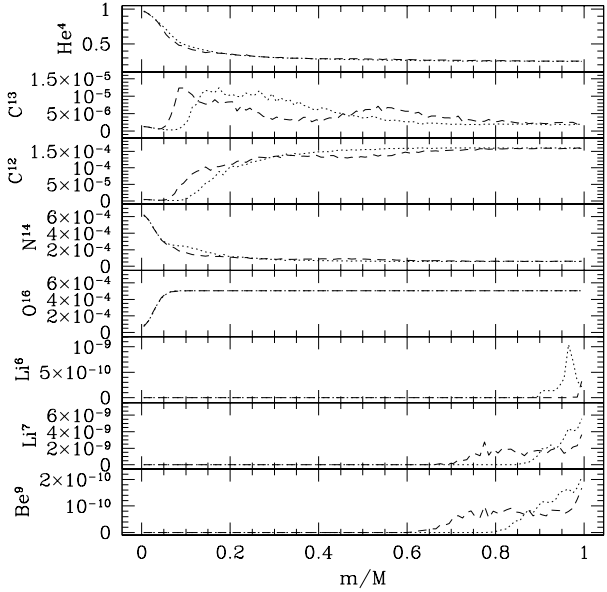}
% chem12
\caption{Chemical abundance profiles for the case 1
(dashed curve) and case 2 (dotted curve) collision products, as
determined by SPH calculations.
The chemical abundance fractions are averaged
on isodensity surfaces that enclose a mass fraction $m/M$ in the final
collision product.}
\label{chem12}
\end{figure}

\begin{figure}
\centering
\includegraphics[width=84mm]{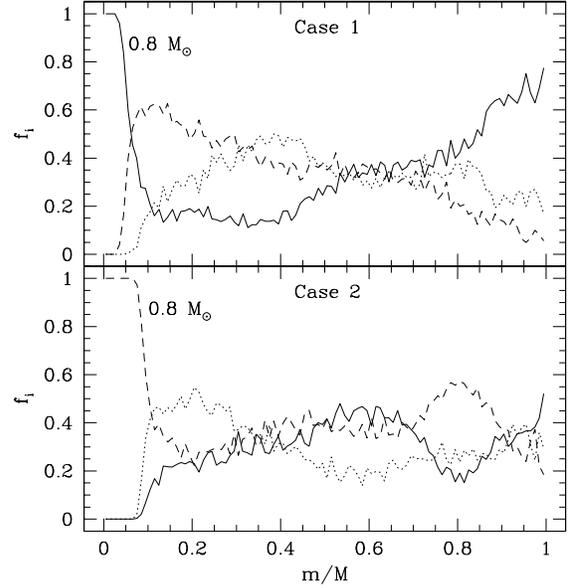}
% frac12:
\caption{Fractional contribution $f_i$ to the mass of the collision product
from each parent star as a
function of enclosed mass fraction $m/M$ within the 
product for case 1 (top) and case 2 (bottom), as determined by SPH
calculations.  Each of these cases involves head-on
($r_p=r_{p,2}=0$) collisions among one $0.8$ and two $0.6 {\rm
 M}_\odot$ stars;
however in case 1 the $0.8 {\rm M}_\odot$ star is part of the initial
collision, whereas in the case 2 scenario it is part of the second
collision.  Different line types
are used for each parent star: $i=1$ (solid curve), $i=2$ (dotted
curve), and $i=3$ (dashed curve).  Parents with an index $i=1$ or 2
are involved in the first collision, while $i=3$ refers to the
third parent star from the second collision.  The contribution profile
from the $0.8 {\rm M}_\odot$ parent is labelled in each case.
}
\label{frac12}
\end{figure}

\begin{figure}
\centering
\includegraphics[width=84mm]{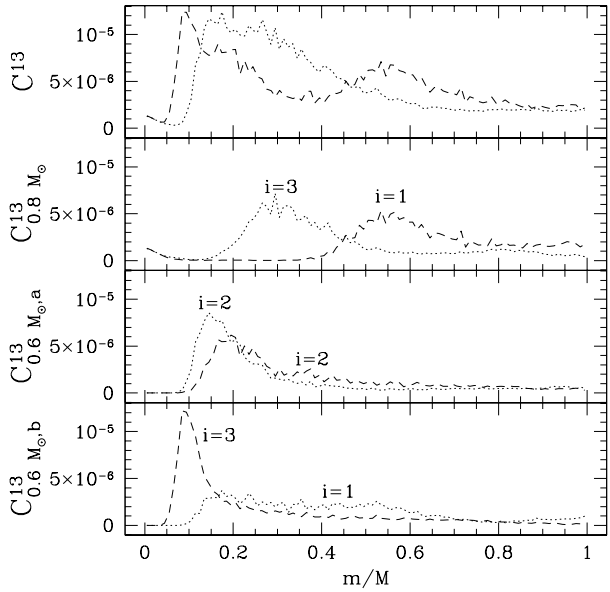}
% f12c13:
\caption{Fractional abundance of C$^{13}$ versus the enclosed mass
fraction $m/M$ in the final collision products of case 1 (dashed curve)
and case 2 (dotted curve), as determined by SPH calculations.
The top pane shows the total C$^{13}$ abundance.  The second pane shows
the contribution from the $0.8 {\rm M}_\odot$ parent, and the bottom
two frames show the contribution from the $0.6 {\rm M}_\odot$ parents,
so that the curves in the bottom three panes add up to give the
overall profile shown in the top pane.
}
\label{f12c13}
\end{figure}

Cases 1 and 2 each involve two consecutive head-on collisions between the
same three parent stars.  The only variation between these
two cases is the order in which the collisions occur:
in case 1, the $0.8 {\rm M}_\odot$ parent is involved in the first
collision, while in case 2 it is involved in the second.
Fig.~\ref{chem12} shows the resulting chemical composition profiles of the collision products.
Because the innermost few per cent of the final collision products
consist of low-$A$ fluid that originated in the centre of the
$0.8{\rm M}_\odot$ parent (see Fig.~\ref{frac12}), the composition profiles are
nearly identical there.  More generally, the resulting profile of each chemical species is at least qualitatively
similar throughout the products.
The differences in the composition profiles of Fig.~\ref{chem12} are
arguably most pronounced for C$^{13}$.  In the parent stars, this element
exists in appreciable amounts only in a relatively thin shell, and,
as shown in Fig.~\ref{refpar2},
this shell is at a higher value of $A$ in the $0.8 {\rm M}_\odot$
star than in the $0.6 {\rm M}_\odot$ star.  Exactly where the C$^{13}$-rich
fluid is ultimately deposited depends on the details
of the shock heating, and hence the order in which the stars collide.
In particular, the final C$^{13}$ profile in the case 1 product has
two distinct peaks at enclosed mass fractions of $m/M \simeq$ 0.1 and
0.5 (as in our zeroth order model, see Fig~\ref{zeroth_order}), whereas
the case 2 profile has a single extended peak centred
near $m/M$ = 0.2.  Fig.~\ref{f12c13} displays, as a function of
enclosed mass fraction $m/M$ within the product, how each parent
contributes to the overall C$^{13}$ profile.
The inner peak in the case 1 profile is due mostly
to C$^{13}$ that originated in the $0.6{\rm M}_\odot$ parent stars, while
the outer peak is due mostly to the higher-$A$, C$^{13}$-rich
fluid from the $0.8{\rm M}_\odot$ parent.  In case 2, the $0.8{\rm M}_\odot$
star is involved in only the second collision.  It therefore
experiences less shock heating than in case 1 and more of its fluid is
able to penetrate to the core of the final collision product.
Consequently, much of the C$^{13}$ from the $0.6 {\rm M}_\odot$ stars is displaced
out to larger enclosed mass fractions, while the C$^{13}$ from the
$0.8 {\rm M}_\odot$ star is shifted inward.  The net result is the single
extended peak that includes C$^{13}$ from all three parent stars.

%%%%%%%%%%%%%%%%%%%%%%%%%%%%%%%%%%%%%%%%%%%%%%%%%%%%%%%%%%%%%
% FIGURE 4
%%%%%%%%%%%%%%%%%%%%%%%%%%%%%%%%%%%%%%%%%%%%%%%%%%%%%%%%%%%%%
%%%%%%%%%%%%%%%%%%%%%%%%%%%%%%%%%%%%%%%%%%%%%%%%%%%%%%%%%%%%%
% FIGURE 5
%%%%%%%%%%%%%%%%%%%%%%%%%%%%%%%%%%%%%%%%%%%%%%%%%%%%%%%%%%%%%
%%%%%%%%%%%%%%%%%%%%%%%%%%%%%%%%%%%%%%%%%%%%%%%%%%%%%%%%%%%%%
% FIGURE 6
%%%%%%%%%%%%%%%%%%%%%%%%%%%%%%%%%%%%%%%%%%%%%%%%%%%%%%%%%%%%%

The detailed differences between the
composition profiles of the other elements in Fig.~\ref{chem12}
can be understood by considering the $A$ and
composition profiles of the parent stars, along with the distribution
and amount of shock heating.  For example, the core of the $0.8
{\rm M}_\odot$ parent star is rich in He$^4$ and N$^{14}$, but depleted of
C$^{12}$ (see Fig.~\ref{refpar2}).  The lower shock heating to the
$0.8 {\rm M}_\odot$ star in case 2
allows more of its core to sink to the centre
of the collision product.  Consequently, the He$^4$ and N$^{14}$ levels
are enhanced as compared to case 1, while the C$^{12}$ levels are diminished, for final
enclosed mass fractions $m/M$ in the range from 0.05 to 0.2.

The mass that is ejected during the collisions comes preferentially
from the outer layers of the parent stars, exactly where
elements such as Li$^6$, Li$^7$, and Be$^9$ exist.  The
surface abundances (by mass) in the final case 1 collision product for
these three elements are approximately $3\times 10^{-10}$, $4\times
10^{-9}$, and $2\times 10^{-10}$, which is, respectively, about 30,
6, and 3 times less than at the surface of the $0.8{\rm M}_\odot$ parent star.  In
case 2, the surface layers are comparably depleted in these elements:
the corresponding abundances are $2\times 10^{-10}$, $6\times 10^{-9}$,
and $2\times 10^{-10}$.

The bottom three panes in Fig.~\ref{chem12} show
that the distribution of Li$^6$, Li$^7$, and Be$^9$ does differ
somewhat between cases 1 and 2.  Because the $0.8 {\rm M}_\odot$ star
suffers less shock heating in case 2 than in case 1, it loses less of
its mass as ejecta and, consequently, can contribute more Li$^6$, Li$^7$,
and Be$^9$ to the outermost layers of the final collision product.
Furthermore, when the fluid containing Li$^7$ and Be$^9$ from the $0.6 {\rm M}_\odot$ stars is
involved in both collisions (case 2), it is shocked more and ultimately either
ejected or deposited in the outer $\sim$10 per cent of the
product. However in case 1, the one $0.6 {\rm M}_\odot$ star that is
involved in only a single collision can deposit its Li$^7$ and Be$^9$
of comparatively low-$A$ deeper in the product, resulting in flattened
profiles extending further into the interior.

Fig.~\ref{frac12} shows, as a function of $m/M$,
the fractional contribution to the final product's mass
from each of the three parent stars for cases 1 and
2, as determined by SPH calculations.
In each case, the innermost few per cent of the final collision product
consists of low-$A$ fluid that originated in the centre of the
$0.8{\rm M}_\odot$ parent.
Because the first collision in cases 1 and 2 is head-on ($r_p=0$),
fluid from the first two parent stars is {\it not} distributed
axisymmetrically in the first
collision product (the composition distribution is therefore not
axisymmetric, even though the structure of the first product is).
In case 2, the $0.8 {\rm M}_\odot$ star
strikes the first collision product on the side with fluid from the
first ($i=1$) $0.6 {\rm M}_\odot$ parent.
Fluid from the first $0.6 {\rm M}_\odot$ parent is therefore heated
more than fluid from the second, and the former 
is buoyed out to larger enclosed mass fractions in the final product.
In off-axis collisions, rotation induces shear mixing, so that if
two identical stars are involved in the first collision, they
contribute essentially equally within the final product: $f_1=f_2$.

\begin{figure}
\centering
\includegraphics[width=84mm]{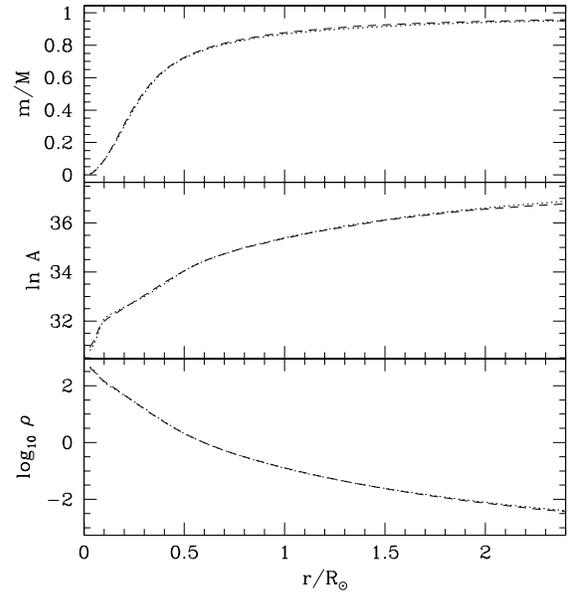}
% slog12:
\caption{Structural profiles for the case 1
(dashed curve) and case 2 (dotted curve) collision products, as
determined by SPH calculations:
the enclosed mass fraction $m/M$,
the natural logarithm of the average entropic variable $A$, and the base 10
logarithm of the average density $\rho$, are all plotted
as a function of the average distance $r$
from the centre of the collision product to an isodensity surface.
Units are cgs.}
\label{slog12}
\end{figure}

The profiles of Fig.~\ref{chem12} and Fig.~\ref{f12c13} 
demonstrate that
the order in which the stars collide can influence shock heating
enough to affect, at least slightly, the chemical composition
distribution within the final collision product.  While the difference
in resulting chemical composition profiles is small, 
Fig.~\ref{slog12} shows that the difference in the structure of the
collision product would be completely negligible for most purposes.
Although changing the order of these head-on
collisions affects how the shock heating is distributed (and hence
where any particular fluid element settles), it does not greatly
affect the overall amount of heating that occurs nor the amount of mass that is ejected.  At least for low Mach number collisions (as with parabolic collisions) that are nearly head-on (so that the merger occurs quickly), shock heating can be thought
of as a mild perturbation; consequently, the final $A$ profile, and hence the
structure of the final product, is not sensitive to the collision order
in such cases.

%%%%%%%%%%%%%%%%%%%%%%%%%%%%%%%%%%%%%%%%%%%%%%%%%%%%%%%%%%%%%
% FIGURE 7
%%%%%%%%%%%%%%%%%%%%%%%%%%%%%%%%%%%%%%%%%%%%%%%%%%%%%%%%%%%%%

% SECTION 3.1.2
\subsubsection{Varying the direction of approach of the third parent \label{direction}}

We now investigate how the direction of approach of the
third star toward the first collision product affects the final
collision product.  One might wonder, for example, whether an impact
in the first collision product's equatorial plane ($\theta=0$ or
180$^\circ$) would yield a qualitatively different result than if the
impact had instead occurred on the rotation axis.  Cases 5 and 6, cases
7 to 10, and cases 14 to 19 can all be used to examine such effects,
as the cases within each set differ only in the angles
$\theta$ and $\phi$, by which the first product is
rotated (see Fig.~\ref{rotate}).
We find that while the spin of the final product is of
course sensitive to such variations (e.g., see the $T/|W|$ column of
Table~\ref{second}), the composition profiles are nearly unaffected.

\begin{figure}
\centering
\includegraphics[width=84mm]{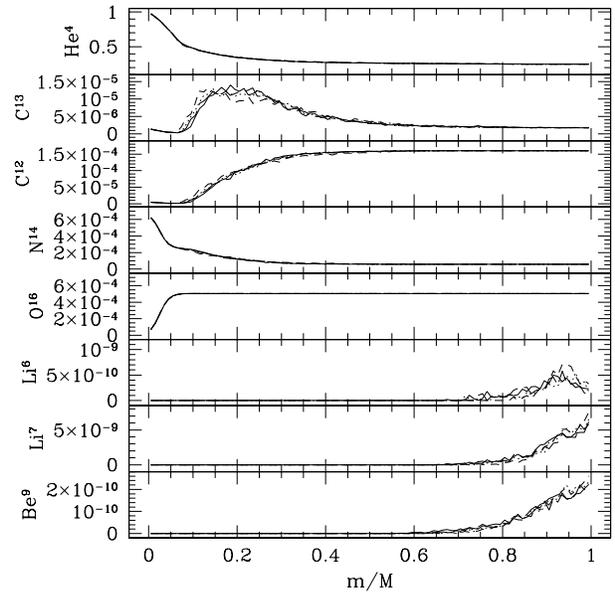}
\caption{Chemical abundance profiles for the final
collision product of cases 14 (solid curve), 15 (dotted curve), 17
(long dashed curve), and 19 (short dashed curve), as determined by SPH
calculations.  In these scenarios the same rotating first
collision product collides off-axis with a $0.8 {\rm M}_\odot$ star,
with a different orientation of the first collision product's rotation
axis in each case.
\label{chem14151719} }
\end{figure}

Fig.~\ref{chem14151719} shows the chemical abundance 
%%%%%%%%%%%%%%%%%%%%%%%%%%%%%%%%%%%%%%%%%%%%%%%%%%%%%%%%%%%%%%%%%
% FIGURE 8
%%%%%%%%%%%%%%%%%%%%%%%%%%%%%%%%%%%%%%%%%%%%%%%%%%%%%%%%%%%%%%%%%
profiles of the collision product resulting in four cases in which the
angle of approach of the third star is varied (cases 14, 15, 17, and
19).  Each of these cases involves an off-axis collision between a
case k collision product and a $0.8 {\rm M}_\odot$ star.  In case 14, the
first collision product's spin vector is parallel to the orbital
angular momentum of the second collision.  In the other cases, the
case k collision product is tilted various ways according to the values
of $\theta$ and $\phi$ listed
in Table \ref{second}.  In case 19, for example, the case k
collision product is flipped over 180$^\circ$ so that it rotates in an
opposite direction to that of the case 14 rotation.

\begin{figure}
\centering
\includegraphics[width=84mm]{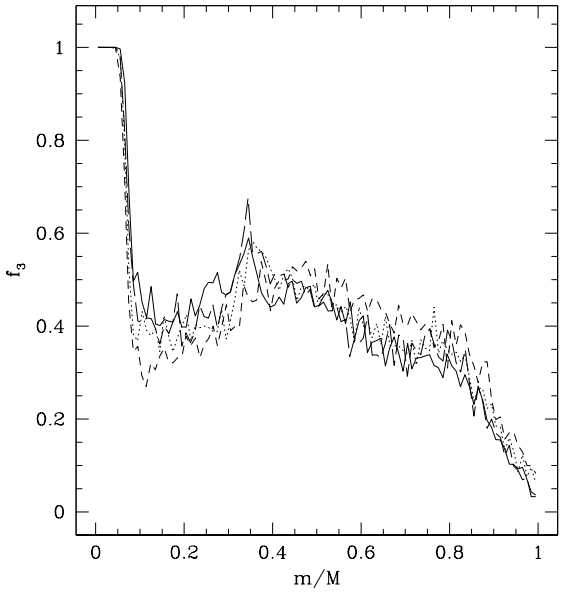}
\caption{
Fractional contribution $f_3$ from the third parent star
as a function of enclosed mass fraction $m/M$ within the collision product
of cases 14 (solid curve), 15 (dotted curve), 17
(long dashed curve), and 19 (short dashed curve), as determined by SPH
calculations.  In these scenarios, the first two parent stars are both $0.6 {\rm M}_\odot$; the third parent star is $0.8 {\rm M}_\odot$ and approaches
from a different angle $\theta$ relative to the rotation of the
first collision product in each case.  The fractional contribution from each of
the first two parent stars is essentially equal and can therefore be
determined easily from the $f_3$ curve: $f_1=f_2=(1-f_3)/2$.
\label{f14151719} }
\end{figure}

In case 14, the fluid of the first product is rotating with the third
star's motion as it impacts ($\theta=0^\circ$).  Consequently, the
merger process is relatively gentle.  For larger $\theta$, the
relative impact velocity is larger and the merger is somewhat more
violent.  Cases 14, 17, 15, and 19 have $\theta$ values
of 0, 45, 90, and 180$^\circ$, respectively; as $\theta$ increases,
slightly less fluid from the $0.8{\rm M}_\odot$ star can sink
down into the core of the final collision product, and the C$^{13}$
profile rises at a slightly smaller enclosed mass fraction $m/M$ (see 
Fig.~\ref{chem14151719}).  Nevertheless, as shown in
Fig.~\ref{f14151719}, the contribution of each
%%%%%%%%%%%%%%%%%%%%%%%%%%%%%%%%%%%%%%%%%%%%%%%%%%%%%%%%%%%%%%%%%
% FIGURE 9
%%%%%%%%%%%%%%%%%%%%%%%%%%%%%%%%%%%%%%%%%%%%%%%%%%%%%%%%%%%%%%%%%
parent star to the product varies very little from case to case.
Consequently, the chemical profiles in
the collision products also vary little as $\theta$ is changed.  Indeed, the
He$^4$, C$^{12}$, N$^{14}$, and O$^{16}$ profiles in Fig.~\ref{chem14151719}
all look remarkably
similar to the corresponding profiles in Fig.~\ref{zeroth_order} for
our simple, zeroth order model.  However, the C$^{13}$ profile
has a single broad peak, for the same reasons
as in case 2.
Furthermore, because of mass loss, the beryllium and lithium surface
abundances are found to be greatly less than our zeroth order model
(that neglects mass loss) would indicate.

\begin{figure}
\centering
\includegraphics[width=84mm]{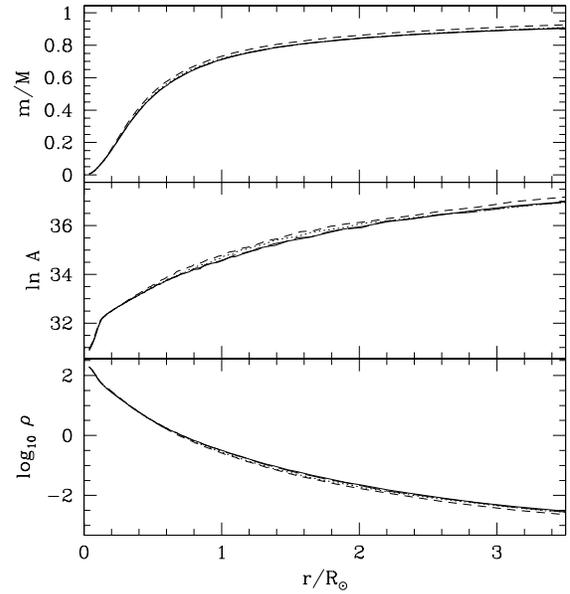}
\caption{Structural profiles for the final
collision product of cases 14 (solid curve), 15 (dotted curve), 17
(long dashed curve), and 19 (short dashed curve), as determined by SPH
calculations.  The particular quantities plotted are as in Fig.~\ref{slog12}.
\label{slog14151719} }
\end{figure}

The structure of the final collision product (see
Fig.~\ref{slog14151719}) can be affected by the direction of approach
%%%%%%%%%%%%%%%%%%%%%%%%%%%%%%%%%%%%%%%%%%%%%%%%%%%%%%%%%%%%%%%%%
% FIGURE 10
%%%%%%%%%%%%%%%%%%%%%%%%%%%%%%%%%%%%%%%%%%%%%%%%%%%%%%%%%%%%%%%%%
for two primary reasons.  Firstly, having larger relative velocity at
impact leads to larger shock heating.  Notice, for example, how the
case 19 product has the largest $A$ values in Fig.~\ref{slog14151719}.
Secondly, having less angular momentum in the system leads to a more
compact product.  For example, the case 19 product has the
largest enclosed mass fraction for almost any average radius $r$,
despite the additional shock heating undergone in this case.
Furthermore, by comparing the final masses $M_{SPH}$ listed in
Table~\ref{second} for the products of cases 5 and 6, of cases 7
through 10, and of cases 14 through 19, one can see that the amount of
mass ejected is only very weakly dependent upon the direction of
approach of the third parent star, varying by about $0.01{\rm M}_\odot$ or
less within each of these sets of cases.

% Section 3.1.3
\subsubsection{Varying the periastron separations of the collisions
\label{spin}}

\begin{figure}
\centering
\includegraphics[width=84mm]{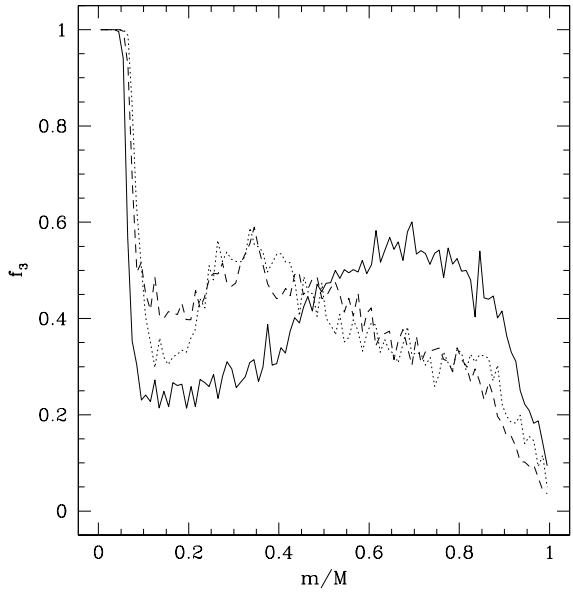}
\caption{Fractional contribution of the third parent star, as
determined by SPH calculations, within the final collision product for
three scenarios that differ only in the periastron separation $r_p$ of
the first collision: cases 12 (solid curve), 13 (dotted curve), and 14
(dashed curve).  In all three cases, the first two parent stars are
$0.6 {\rm M}_\odot$, while the third parent is $0.8 {\rm M}_\odot$.
The $0.8 {\rm M}_\odot$ parent penetrates into the product the least
in case 12, because of the relatively small amount of shock heating
suffered by the first product during the first collision in this case.
\label{frac121314}}
\end{figure}

We now investigate the effects that the periastron separation of the
first collision has on the final collision product.  Cases 12, 13, and
14 all involve off-axis collisions with first collision products that
are created from the same $0.6 {\rm M}_\odot$ parent stars but with
different periastron separations (cases j, jk, and k, respectively),
and hence different inherited angular momenta.
Fig.~\ref{frac121314} plots the fractional contribution of the third
parent star within the merger product.
%%%%%%%%%%%%%%%%%%%%%%%%%%%%%%%%%%%%%%%%%%%%%%%%%%%%%%%%%%%%%%%%%
% FIGURE 11
%%%%%%%%%%%%%%%%%%%%%%%%%%%%%%%%%%%%%%%%%%%%%%%%%%%%%%%%%%%%%%%%%
In all cases, the low-$A$ core of the $0.8{\rm M}_\odot$ star
is able to sink to the core of the final product.  However,
as the periastron separation of the first collision is increased,
the two $0.6 {\rm M}_\odot$ parent stars experience more shock
heating, and the $0.8 {\rm M}_\odot$ parent is able to have more fluid
penetrate down to the depths near $m/M\sim 0.15$.

\begin{figure}
\centering
\includegraphics[width=84mm]{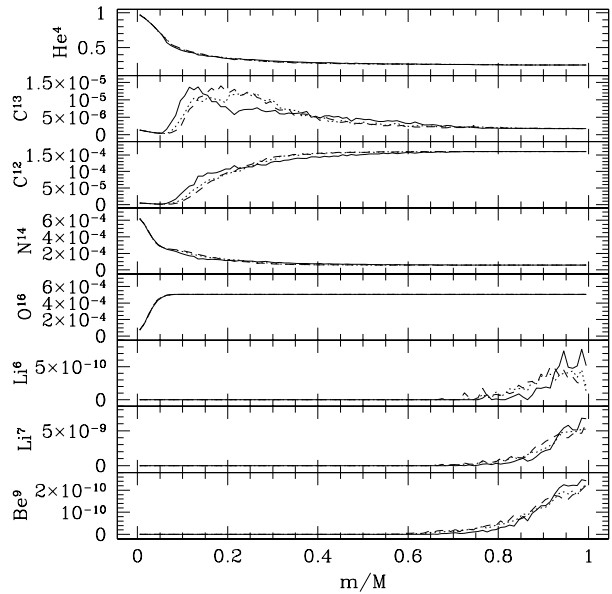}
\caption{Chemical
abundance profiles, as determined by SPH
calculations, for the final collision product
for three scenarios that differ only in the periastron separation $r_p$ of the first collision: cases 12 (solid
curve), 13 (dotted curve), and 14 (dashed curve).
\label{chem121314}}
\end{figure}

\begin{figure}
\centering
\includegraphics[width=84mm]{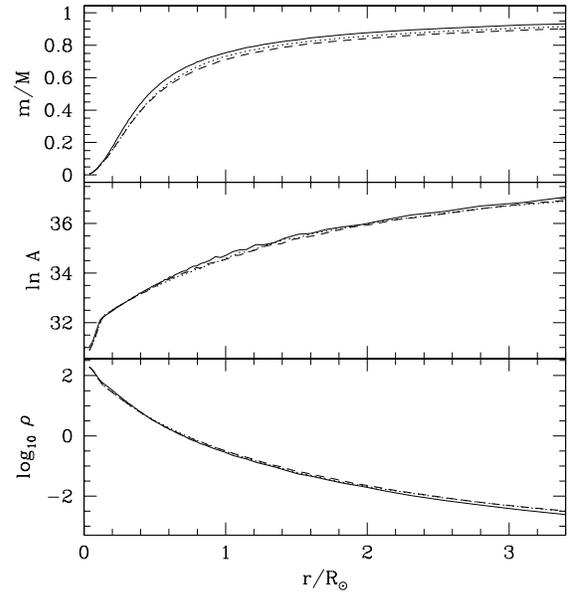}
\caption{Structural profiles, as determined by SPH
calculations, for the final collision product
for three scenarios that differ only in the periastron separation $r_p$ of the first collision: cases 12 (solid
curve), 13 (dotted curve), and 14 (dashed curve).  The particular
quantities plotted are as in Fig.~\ref{slog12}.
\label{slog121314}}
\end{figure}

Fig.~\ref{chem121314} presents chemical composition profiles
%%%%%%%%%%%%%%%%%%%%%%%%%%%%%%%%%%%%%%%%%%%%%%%%%%%%%%%%%%%%%%%%%
% FIGURE 12
%%%%%%%%%%%%%%%%%%%%%%%%%%%%%%%%%%%%%%%%%%%%%%%%%%%%%%%%%%%%%%%%%
of the final collision product for these three cases.
These profiles demonstrate that the angular momentum of the first
collision product has only a small effect on the final collision
product.  As expected from Fig.~\ref{frac121314}, the profiles
of the three cases are essentially identical in the innermost 5 per
cent of the bound mass, because only the core of the $0.8\
{\rm M}_\odot$ parent contributes there.
The abundance profiles of each chemical species are at least
qualitatively, and usually quantitatively, similar throughout the
products.  The variations that do exist can be
understood in terms of the different shock heating during the first
collisions.  Because the amount of shock heating increases with
periastron separation, the case j, jk, and k products have
increasingly larger values of $A$ at almost any enclosed mass
fraction (this trend is not immediately evident in Fig.~\ref{slog121314}
only because $\ln A$ is being plotted versus radius and not enclosed mass).
The fluid from the $0.8{\rm M}_\odot$ star is therefore able
to penetrate the case j product the least, the case jk product a
little more, and the case k product even more still (see Fig.~\ref{frac121314}).  Consequently the
rise in C$^{12}$ and C$^{13}$ abundance is pushed out to increasingly
larger enclosed mass fractions $m/M$ in Fig.~\ref{chem121314} as one
considers cases 12, 13, and 14, in that order.  In case 12 and
arguably case 13, the cases with the lesser amounts of shock heating,
traces of two separate peaks are evident in the C$^{13}$ profile.  As
in our zeroth order model (see Fig.~\ref{zeroth_order}), the inner
peak is due mostly to C$^{13}$ from the $0.6{\rm M}_\odot$ stars while the
outer peak is mostly due to C$^{13}$ from the $0.8{\rm M}_\odot$ star.

%%%%%%%%%%%%%%%%%%%%%%%%%%%%%%%%%%%%%%%%%%%%%%%%%%%%%%%%%%%%%%%%%
% FIGURE 13
%%%%%%%%%%%%%%%%%%%%%%%%%%%%%%%%%%%%%%%%%%%%%%%%%%%%%%%%%%%%%%%%%

Fig.~\ref{slog121314} shows that the structure of
the bulk of the fluid in the final collision product is not
significantly affected by the periastron separation of the first
collision, and hence the spin of the first collision product.  There
is, nevertheless, a visible trend for the enclosed mass fraction at a given
average radius to
decrease for products with more spin.  For example, the isodensity
surface with an average radius of $3.2{\rm R}_\odot$ encloses about 94 per
cent of the case 12 product, about 92 per cent for the case 13
product, and only about 90 per cent of the case 14 product.  Such a
trend is expected, simply because of expansion due to rotational support.

Cases 10, 17, and 20 can be used to 
investigate the effects that the periastron
separation of the second collision has on the
profiles of the final product.
Cases 10, 17, and 20
involve collisions between a case k product and a $0.8{\rm M}_\odot$ star,
with periastron separations for the second collision of $r_{p,2}=0$,
0.505, and $0.758{\rm R}_\odot$, respectively, corresponding to a number of
passages or interactions $n_p=1$, 2, and 3, again respectively.
See the discussion of fig.~6 in \citet{lom96} for the details of how
$n_p$ is determined.

\begin{figure}
\centering
\includegraphics[width=84mm]{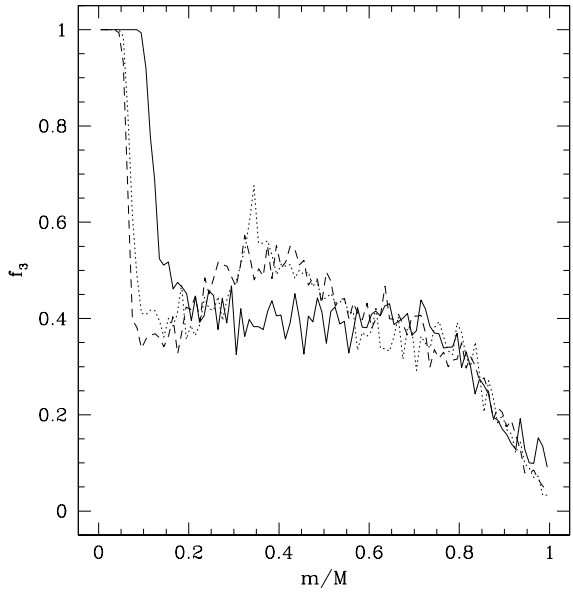}
\caption{Fractional contribution of the third parent star, as
determined by SPH calculations, within the final collision product for
three scenarios that differ only in the periastron separation $r_{p,2}$ of
the second collision: cases 10 (solid curve), 17 (dotted curve), and 20
(dashed curve).  In all three cases, the first two parent stars are
$0.6 {\rm M}_\odot$, while the third parent is $0.8 {\rm M}_\odot$.
The first collision product results from case k.
\label{frac101720}}
\end{figure}

Fig.~\ref{frac101720} reveals the way in which the $0.8 {\rm M}_\odot$ parent
%%%%%%%%%%%%%%%%%%%%%%%%%%%%%%%%%%%%%%%%%%%%%%%%%%%%%%%%%%%%%%%%%
% FIGURE 14
%%%%%%%%%%%%%%%%%%%%%%%%%%%%%%%%%%%%%%%%%%%%%%%%%%%%%%%%%%%%%%%%%
contributes to the final product in each of these three cases.
As usual, the low-$A$
core of the $0.8{\rm M}_\odot$ star sinks to the core of the collision
product.  As the periastron separation of the second collision is
increased, the resulting collision products tend toward larger
mass-averaged values of $A$.
Fluid from the
the $0.8{\rm M}_\odot$ star therefore can penetrate the case 10
product the most, the case 17 product a little less, and the case 20
product even less still.

\begin{figure}
\centering
\includegraphics[width=84mm]{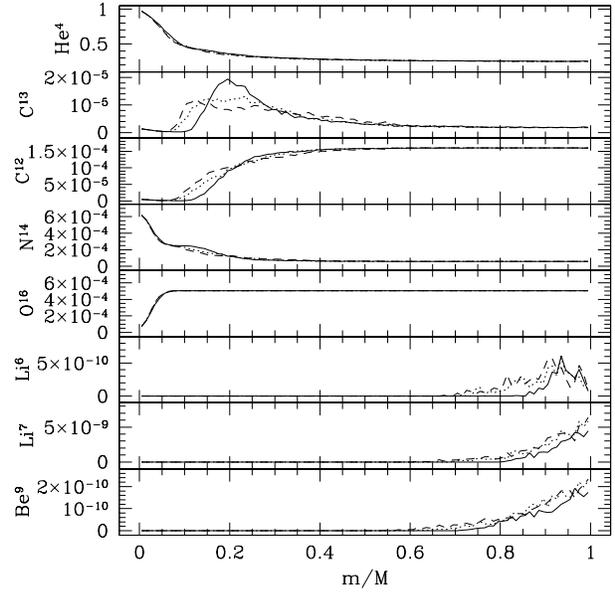}
\caption{Chemical
abundance profiles, as determined by SPH
calculations, for the final collision product
for three scenarios that differ only in the periastron separation $r_{p,2}$ of the second collision: cases 10 (solid
curve), 17 (dotted curve), and 20 (dashed curve).
\label{chem101720}}
\end{figure}

\begin{figure}
\centering
\includegraphics[width=84mm]{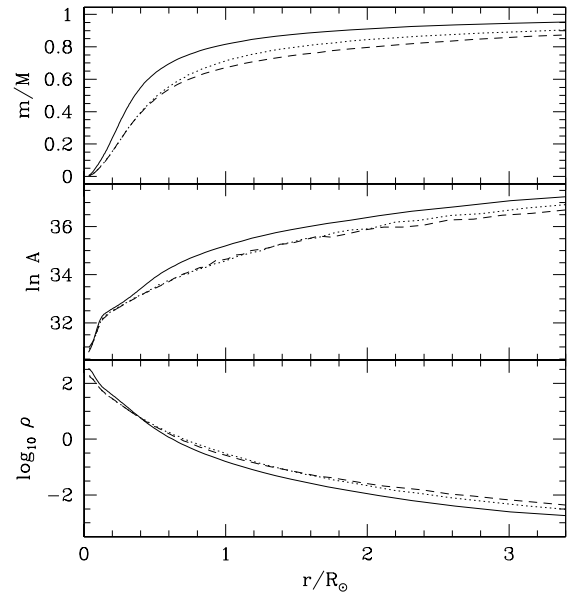}
\caption{Structural profiles, as determined by SPH
calculations, for the final collision product
for three scenarios that differ only in the periastron separation $r_{p,2}$ of the second collision: cases 10 (solid
curve), 17 (dotted curve), and 20 (dashed curve).
\label{slog101720}}
\end{figure}

Fig.~\ref{chem101720} shows that the composition
profiles in these three cases are essentially identical
in the innermost few per cent of the collision products.  Indeed, the
abundance profiles are again at least qualitatively, and usually
quantitatively, similar throughout the entire product.  As before,
slight variations do result from having different distributions of
shock heating.  In particular, the rise in C$^{12}$ and
C$^{13}$ abundance is drawn in to smaller enclosed mass fractions
$m/M$ in Fig.~\ref{chem101720} as one examines cases 10, 17, and 20, in
that order.  Fig.~\ref{slog101720} reveals the differences in the
final product structure for these three cases.
The top pane shows that the
mass distribution of the final product is affected by the periastron
separation of the second collision in a way that is simple to
understand: increasing the second periastron separation increases both
shock heating and rotation, and so a given radius encloses less mass.
%%%%%%%%%%%%%%%%%%%%%%%%%%%%%%%%%%%%%%%%%%%%%%%%%%%%%%%%%%%%%%%%%
% FIGURE 15
%%%%%%%%%%%%%%%%%%%%%%%%%%%%%%%%%%%%%%%%%%%%%%%%%%%%%%%%%%%%%%%%%
%%%%%%%%%%%%%%%%%%%%%%%%%%%%%%%%%%%%%%%%%%%%%%%%%%%%%%%%%%%%%%%%%
% FIGURE 16
%%%%%%%%%%%%%%%%%%%%%%%%%%%%%%%%%%%%%%%%%%%%%%%%%%%%%%%%%%%%%%%%%

%Section 3.2
\subsection{Fluid sorting with {\sevensize MMAS}\label{mmasresults}}

% Section 3.2.1
\subsubsection{Comparison with SPH results \label{comparison}}

%%%%%%%%%%%%%%%%%%%%%%%%%%%%%%%%%%%%%%%%%%%%%%%%%%%%%%%%%%%%%%%
% FIGURE 17
\begin{figure*}
\centering
\includegraphics[width=84mm]{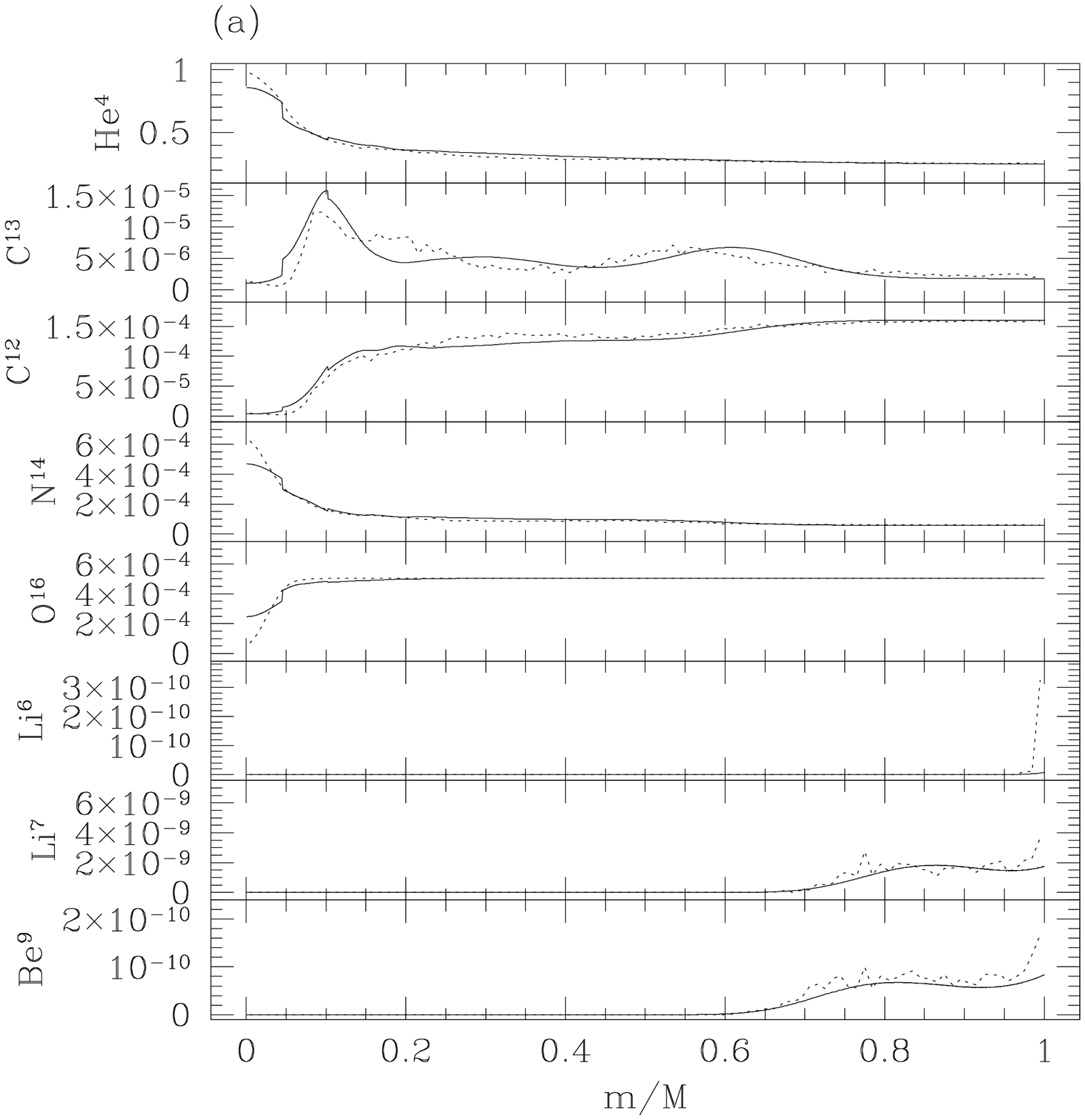}
\includegraphics[width=84mm]{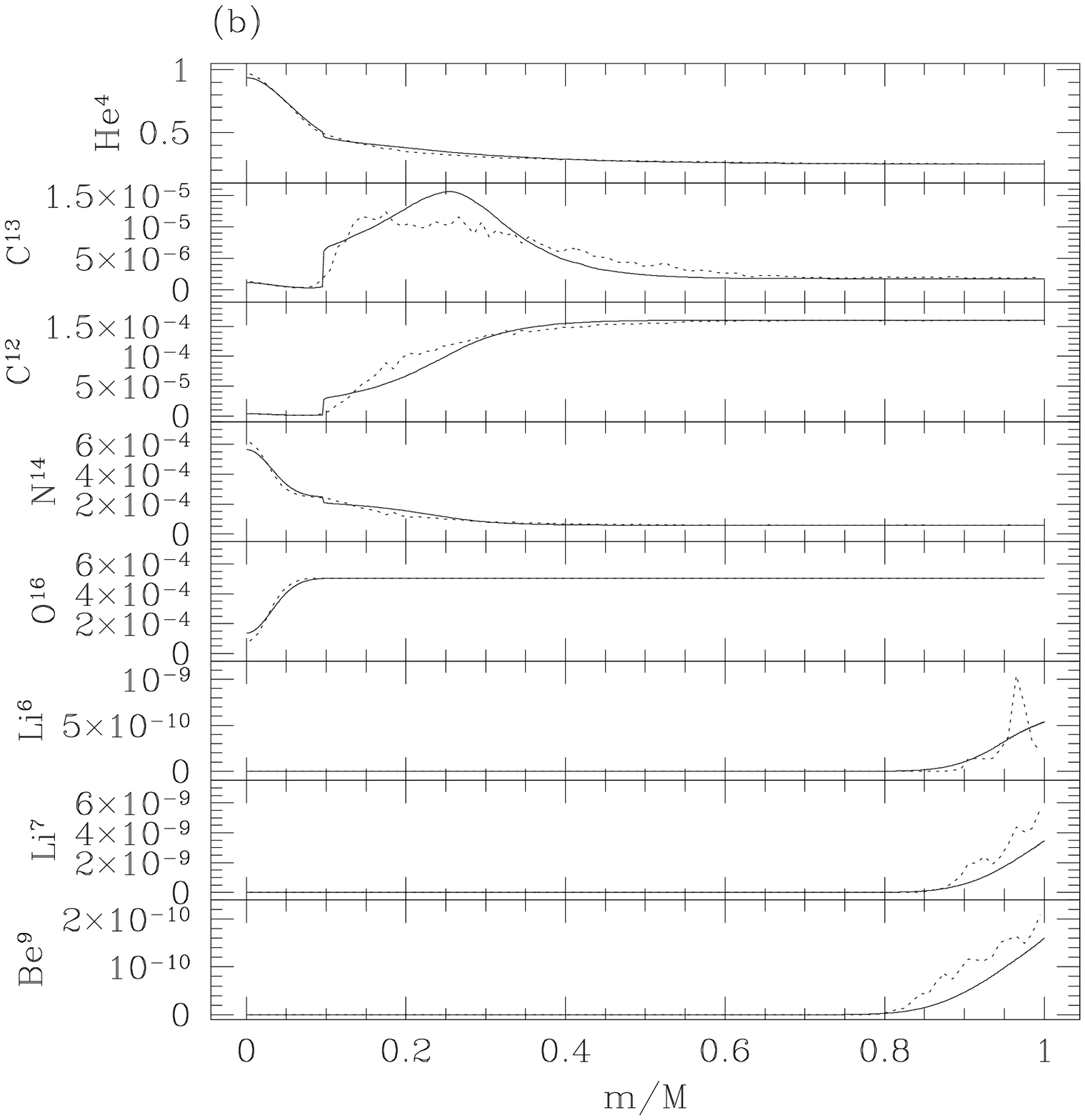}
%chem1mmas,chem2mmas
\caption{Chemical abundance fractions by mass vs.\ the
enclosed mass fraction $m/M$ in the final collision product for (a)
case 1 and (b) case 2.  Results both from an SPH simulation (dotted
curve) and from the {\sevensize MMAS} software package (solid curve) are shown.
In each case, both collisions are head-on.
\label{chem12mmas} }
\end{figure*}
%%%%%%%%%%%%%%%%%%%%%%%%%%%%%%%%%%%%%%%%%%%%%%%%%%%%%%%%%%%%%%%
%%%%%%%%%%%%%%%%%%%%%%%%%%%%%%%%%%%%%%%%%%%%%%%%%%%%%%%%%%%%%%%
% FIGURE 18
\begin{figure*}
\centering
\includegraphics[width=84mm]{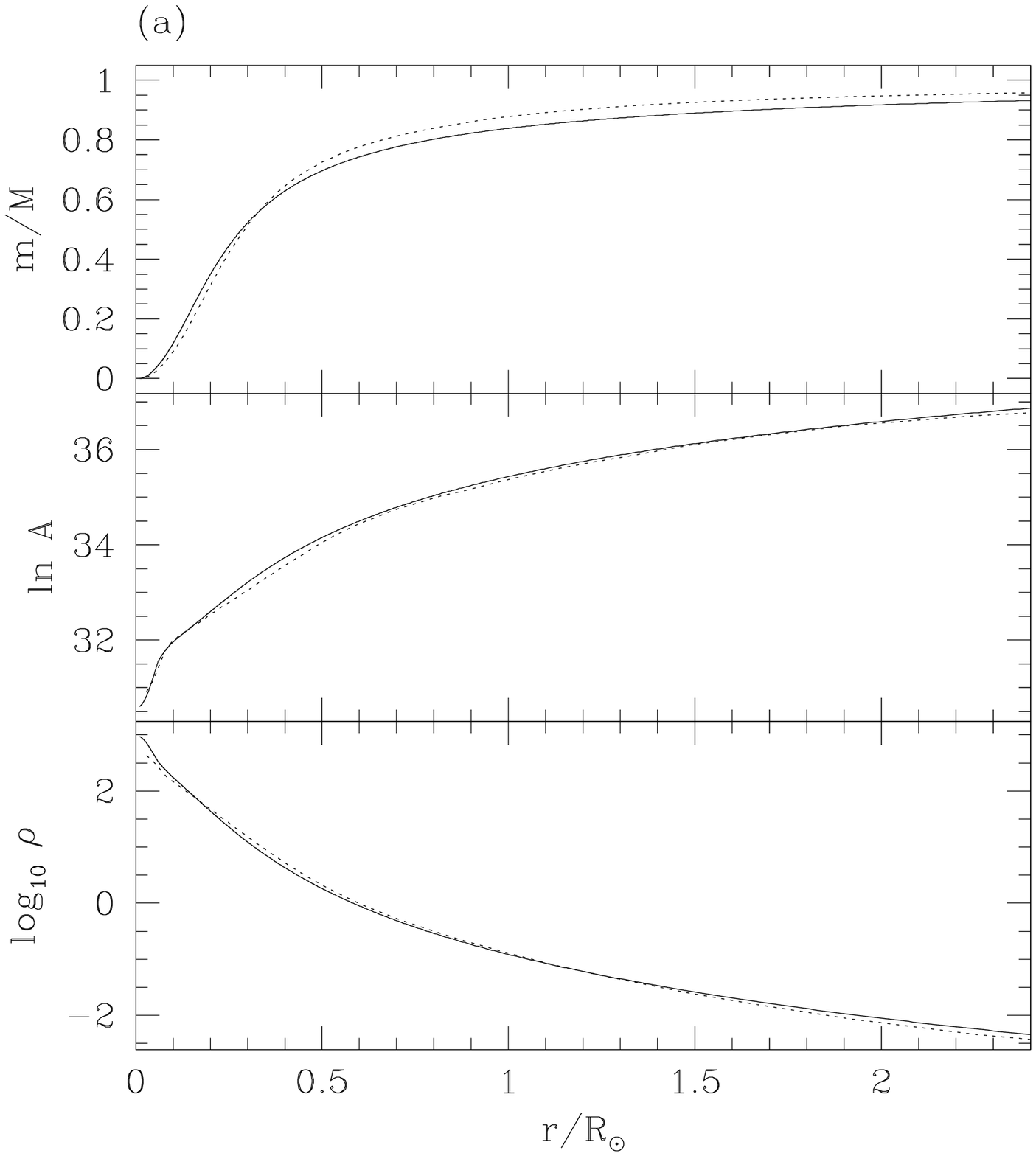}
\includegraphics[width=84mm]{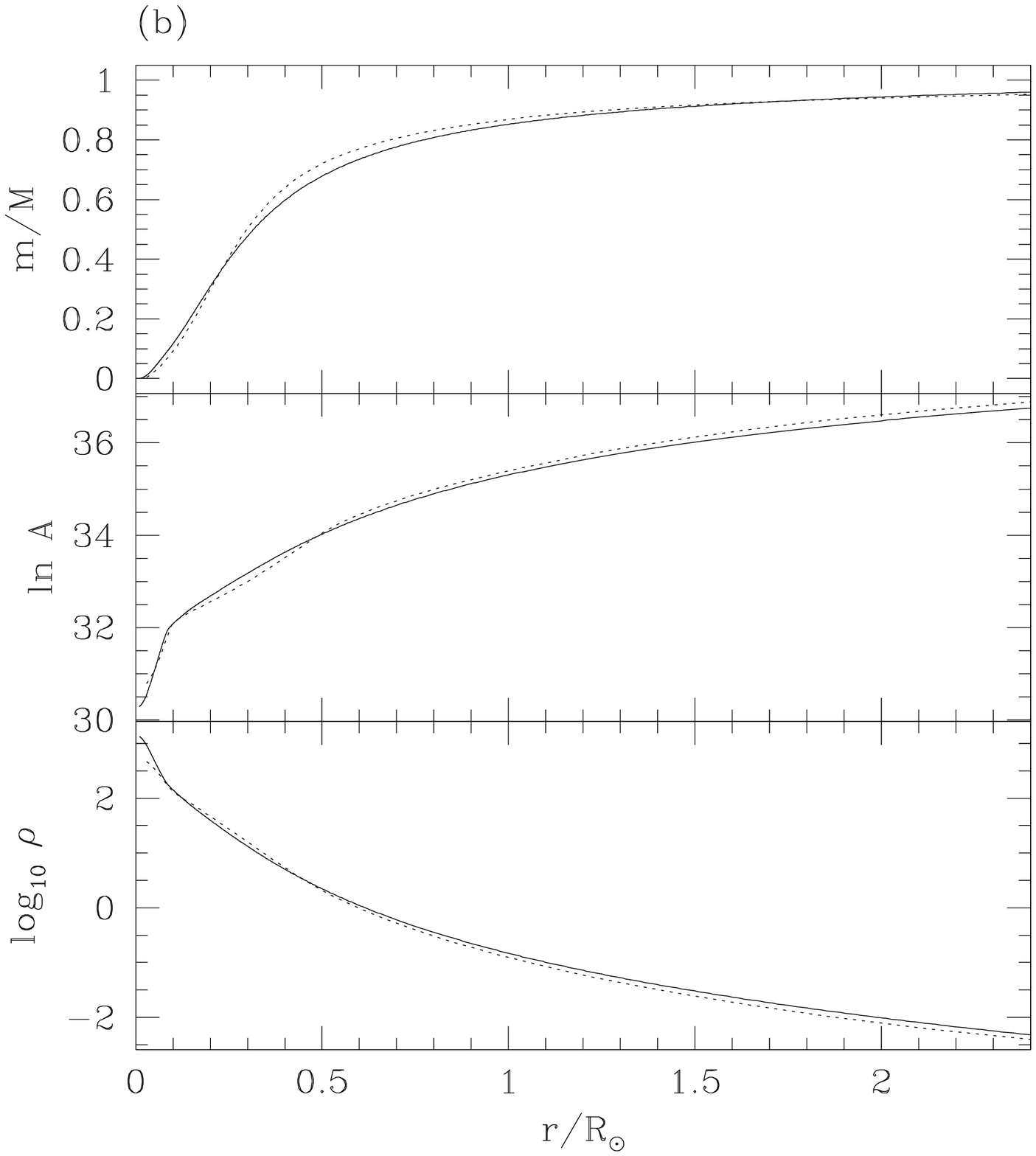}
\caption{Structure profiles as a function of radius $r$ in the final
collision product for (a) case 1 and (b) case 2.  Results both from an
SPH simulation (dotted curve) and from the {\sevensize MMAS} software
package (solid curve) are shown.  In each case, the final collision
product is non-rotating.
\label{slog12mmas} }
\end{figure*}
%%%%%%%%%%%%%%%%%%%%%%%%%%%%%%%%%%%%%%%%%%%%%%%%%%%%%%%%%%%%%%%

In \S\ref{direction} we found that the direction of approach of the
third star only weakly affects the profiles and mass of the final
product. We therefore do not account for the angles $\theta$
and $\phi$ of the second collision when applying our fluid sorting
package {\sevensize MMAS}.  As
a result, the product model
that {\sevensize MMAS} generates
is identical within each of the following sets: cases 5 and 6,
cases 7 through 10, and cases 14 through 19.  In all twenty cases
presented in Table \ref{second}, the final product mass given by
{\sevensize MMAS} agree with those from SPH to within 1.5 per cent.

Fig.~\ref{chem12mmas} compares the chemical composition profiles of
final collision products, as determined by both {\sevensize MMAS} and
SPH models, for two scenarios (cases 1 and 2) in which each collision
is head-on ($r_p=r_{p,2}=0$).  These cases involve the same three parent
stars; however the order in which the stars collide is varied.  The
{\sevensize MMAS} abundance profiles maintain the same qualitative
shape as those of the SPH data for almost all of the elements.  One
possibly important difference is that the {\sevensize MMAS} package slightly
over-mixes the core in case 1, and, consequently, the central helium
abundance is not quite as high as in the SPH calculation.  Another
noteworthy difference is that the Li$^6$ profile, especially in case 1, is
not well represented near the surface.
Because Li$^6$ exists in an
even thinner shell
at the surface of the $0.8{\rm M}_\odot$ star
than does Li$^7$ and Be$^9$, its abundance profile
in the product is particularly sensitive to the mass loss distribution during
the collisions.  Note that {\sevensize
MMAS} does correctly predict that most Li$^6$ is ejected during the
collisions.  Furthermore, the abundance profiles generated by {\sevensize MMAS}
much more closely resemble the SPH results than our zeroth order model does
(see Fig.~\ref{zeroth_order}), indicating that {\sevensize MMAS} is capturing
the important effects of mass loss and shock heating.

In scenarios such as cases 1 and 2 for which the final product is
non-rotating, it is straightforward to obtain the enclosed mass $m$
and density $\rho$ profiles from the $A$ profile by integrating the
equation of hydrostatic equilibrium (see \S\ref{sorting}).
Fig.~\ref{slog12mmas} shows the resulting structure of the final
collision products.  The kink in the $A$ profile a little inside
$r=0.1{\rm R}_\odot$ marks the boundary within which fluid from only the
$0.8{\rm M}_\odot$ star contributes, and {\sevensize MMAS} reproduces this
feature quite well.  The central density of the SPH
model is slightly less than that of the {\sevensize MMAS} model,
mostly due to how density is
calculated as a smoothed average in SPH.
Despite this difference,
the overall structure of
the collision product is extremely well reproduced by {\sevensize MMAS}.

\begin{figure}
\centering
\includegraphics[width=84mm]{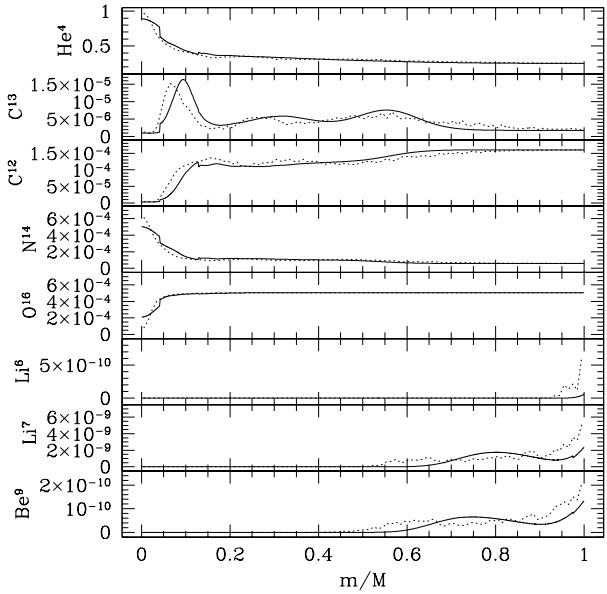}
%chem4mmas
\caption{Chemical abundance fraction by mass versus the enclosed mass
fraction $m/M$ in the final collision product for both the {\sevensize
MMAS} (solid curve) and the SPH (dotted curve) data of the case 4 collision product.
\label{chem4mmas} }
\end{figure}

\begin{figure}
\centering
\includegraphics[width=84mm]{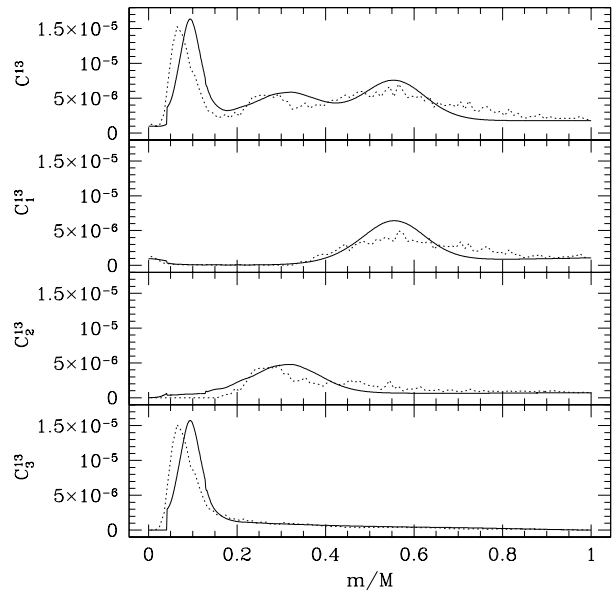}
%compare4c13
\caption{Fractional abundance of C$^{13}$ versus the enclosed mass
fraction $m/M$ in the final case 4 collision product, as determined
both by {\sevensize
MMAS} (solid curve) and by SPH (dotted curve).
The top pane shows the total C$^{13}$ abundance, while the bottom
three panes show the contributions from each individual parent star:
C$^{13}_1$ is the contribution from the $0.8 {\rm M}_\odot$ parent,
C$^{13}_2$ is the contribution from the $0.6 {\rm M}_\odot$ parent in
the first collision, and
C$^{13}_3$ is the contribution from the $0.6 {\rm M}_\odot$ parent in
the second collision.
\label{compare4c13} }
\end{figure}

Fig.~\ref{chem4mmas} compares the chemical composition profiles for
SPH and {\sevensize MMAS} data for case 4,
a situation in which both collisions are off-axis.  The most noticeable
discrepancies are that {\sevensize MMAS} again slightly over-mixes the core
and underestimates the surface Li$^6$ abundance.  Nevertheless,
the chemical abundance profiles produced by the {\sevensize
MMAS} package and the SPH code are extremely similar.
For example, {\sevensize MMAS} correctly reproduces the C$^{13}$
abundance, with three peaks each corresponding to a
different parent star.  The inner peak is due to the low-$A$ fluid
from the $0.6{\rm M}_\odot$ star involved in only the second collision, the
middle peak represents fluid from the other $0.6{\rm M}_\odot$ star, and the
outer peak represents high-$A$ fluid from the $0.8{\rm M}_\odot$ parent
(see Fig.~\ref{compare4c13}).
%%%%%%%%%%%%%%%%%%%%%%%%%%%%%%%%%%%%%%%%%%%%%%%%%%%%%%%%%%%%%%%
% FIGURE 19
%%%%%%%%%%%%%%%%%%%%%%%%%%%%%%%%%%%%%%%%%%%%%%%%%%%%%%%%%%%%%%%
%%%%%%%%%%%%%%%%%%%%%%%%%%%%%%%%%%%%%%%%%%%%%%%%%%%%%%%%%%%%%%%
% FIGURE 20
%%%%%%%%%%%%%%%%%%%%%%%%%%%%%%%%%%%%%%%%%%%%%%%%%%%%%%%%%%%%%%%
Note that this feature is reproducible by {\sevensize MMAS} only
because it accounts for shock heating in each collision (compare to
our zeroth order model of Fig.~\ref{zeroth_order}, in which there are
only two peaks in the C$^{13}$ profile).

\begin{figure}
\centering
\includegraphics[width=84mm]{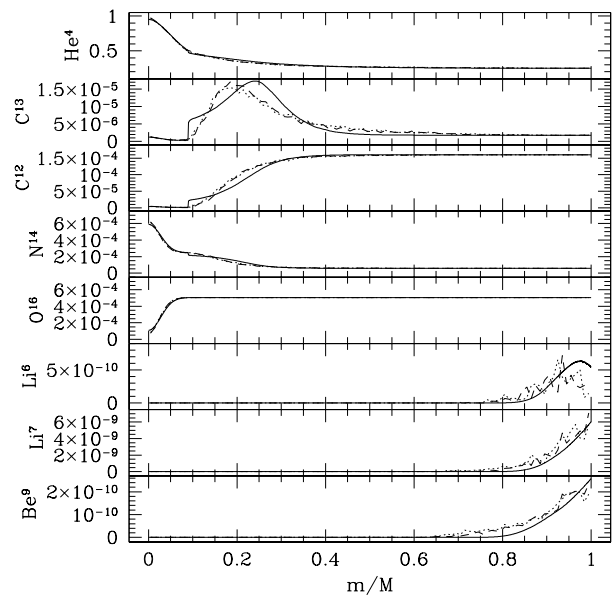}
%chem56mmas
\caption{Chemical abundance fraction by mass versus the enclosed mass
fraction $m/M$ in the final collision product for both the {\sevensize
MMAS} (solid curve) and the SPH data of the 
case 5 (dotted curve) and case 6 (dashed curve) collision products.
Note that our {\sevensize MMAS} results
do not distinguish between cases 5 and 6, as they differ only in the
orientation of the first product's spin.
\label{chem56mmas} }
\end{figure}

\begin{figure}
\centering
\includegraphics[width=84mm]{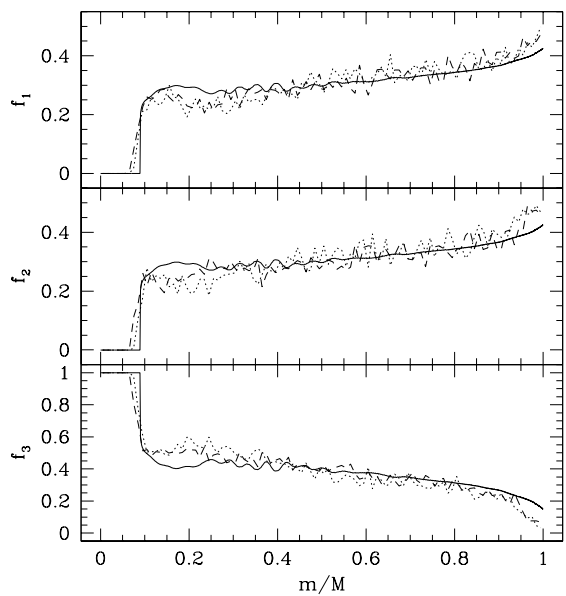}
%compare56
\caption{Fractional contribution $f_i$ of each parent star $i$
versus the enclosed mass
fraction $m/M$ in the case 5 and 6 collision products, as determined
both by {\sevensize
MMAS} (solid curve) and by SPH.  The dotted curve corresponds to the
case 5 SPH results, while the dashed curve gives the case 6 SPH
results.  The same {\sevensize MMAS} model
is valid for both cases, as they differ only in the direction of the
first product's rotation axis.  The $i=1$ and 2 parents are $0.6 {\rm
  M}_\odot$, while the $i=3$ parent is $0.8 {\rm M}_\odot$.
\label{compare56} }
\end{figure}

In many of the {\sevensize MMAS} models, small kinks, or
 discontinuities, are evident in some
of the abundance profiles: such features
mark locations outside of which an additional parent star either
 starts or stops contributing.
For example, in the case 5 and 6 collision product,
a kink exists in the C$^{12}$ and C$^{13}$ profiles near $m/M=0.08$
(see Fig.~\ref{chem56mmas}).
As is evident from Fig.~\ref{compare56}, fluid inside of the
%%%%%%%%%%%%%%%%%%%%%%%%%%%%%%%%%%%%%%%%%%%%%%%%%%%%%%%%%%%%%%%
% FIGURE 21
%%%%%%%%%%%%%%%%%%%%%%%%%%%%%%%%%%%%%%%%%%%%%%%%%%%%%%%%%%%%%%%
%%%%%%%%%%%%%%%%%%%%%%%%%%%%%%%%%%%%%%%%%%%%%%%%%%%%%%%%%%%%%%%
% FIGURE 22
%%%%%%%%%%%%%%%%%%%%%%%%%%%%%%%%%%%%%%%%%%%%%%%%%%%%%%%%%%%%%%%
$m/M\approx0.08$ shell originated
solely in the $0.8{\rm M}_\odot$ parent star. In the range $m/M \ga 0.08$,
all three parent stars contribute.
The smoothing that is inherent to the
SPH scheme makes it difficult to resolve such features with our
hydrodynamics code.  It is possible
that similarly abrupt changes in abundance could occur in nature
within real collision products.

By comparing the SPH data within Fig.~\ref{chem56mmas},
as well as within Fig.~\ref{compare56},
we also see an example of the trend discussed in \S \ref{direction}.
Namely, the direction of the first collision product's rotation axis
(or equivalently the direction of approach of the third parent)
has little effect on the final
collision product.  Indeed, when using {\sevensize MMAS}, our approach
is to neglect completely the rotation of the first collision product,
which is why the same {\sevensize MMAS} model applies to both cases 5
and 6.

%%%%%%%%%%%%%%%%%%%%%%%%%%%%%%%%%%%%%%%%%%%%%%%%%%%%%%%%%%%%%%%
% FIGURE 23
\begin{figure*}
\centering
\includegraphics[width=84mm]{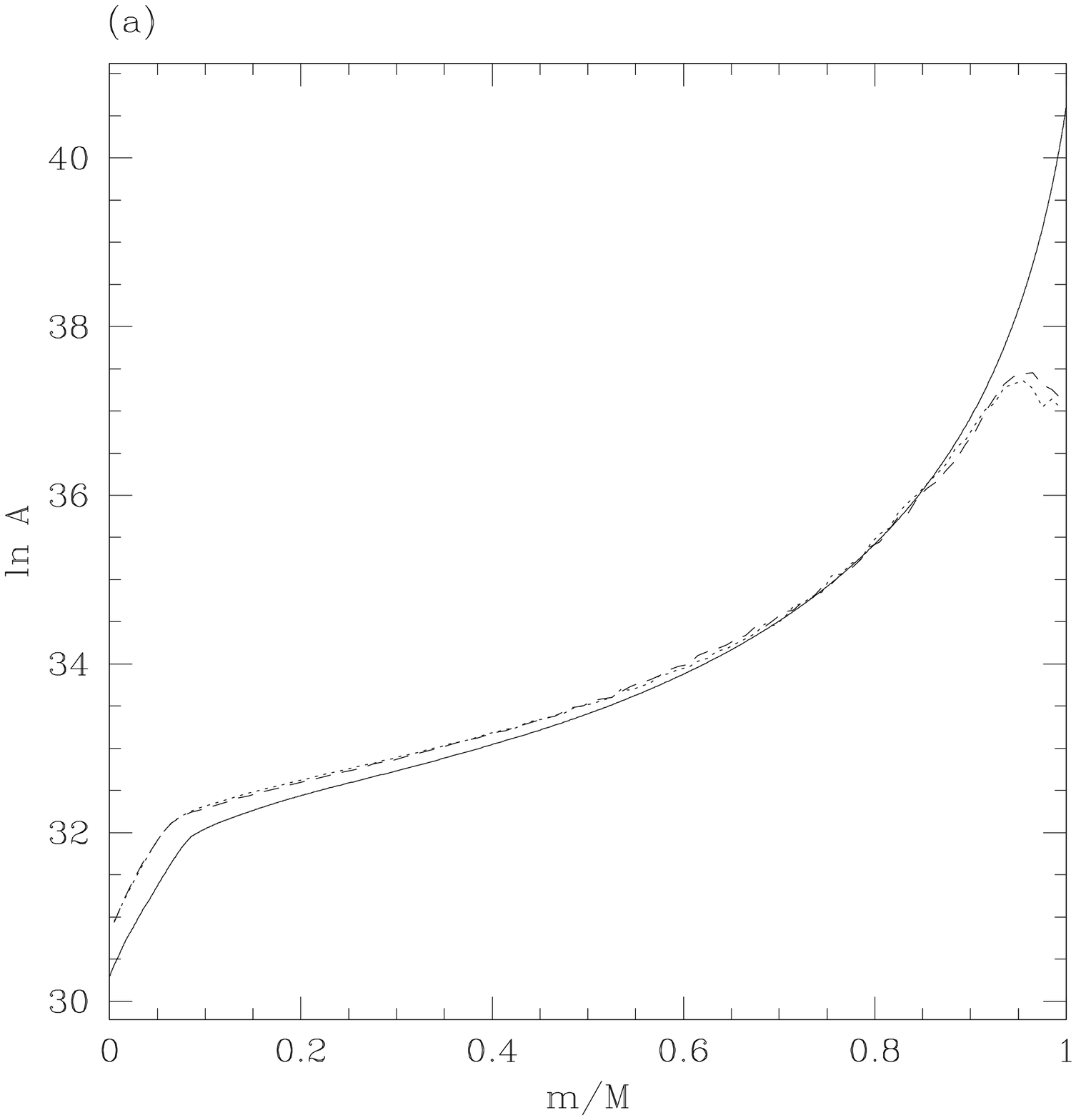}
\includegraphics[width=84mm]{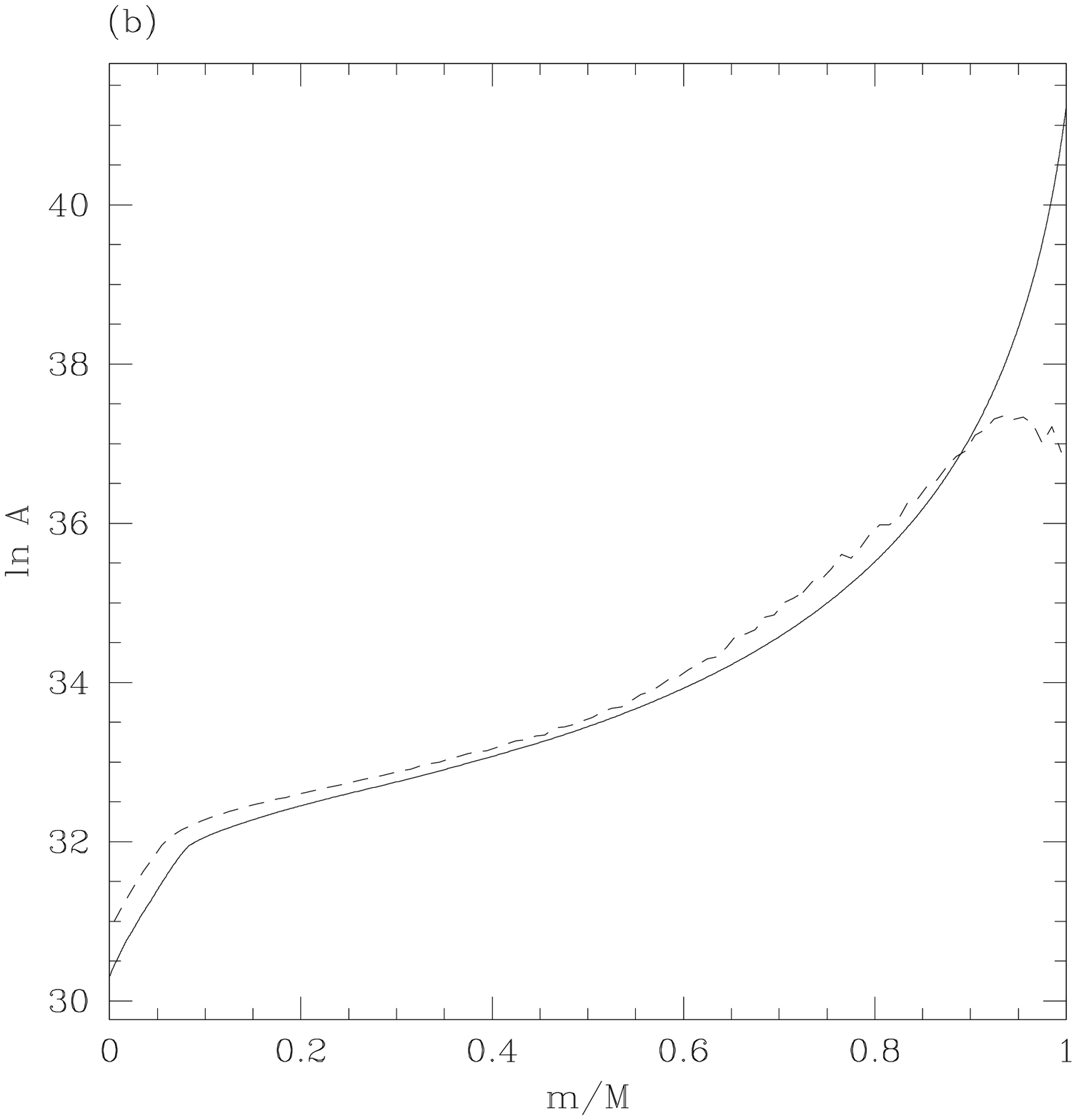}
\caption{Entropic variable $A$ as a function of enclosed mass fraction
$m/M$ in representative final collision products.  In frame (a), the
same {\sevensize MMAS} model (solid curve) is compared against our SPH
models for cases 18 (dotted curve) and 19 (dashed curve): our
implementation of {\sevensize MMAS} does not distinguish between these
cases, as they differ only in the orientation of the first collision
product's rotation axis.  In frame (b), the {\sevensize MMAS} (solid
curve) and SPH (dashed curve) results are compared for case 20.  The
{\sevensize MMAS} profiles have the same qualitative form as the SPH
results, except in the outer $\sim 10$ per cent of the bound mass
where the SPH models have not settled into equilibrium.
\label{slog1819mmas} }
\end{figure*}
%%%%%%%%%%%%%%%%%%%%%%%%%%%%%%%%%%%%%%%%%%%%%%%%%%%%%%%%%%%%%%%

Fig.~\ref{slog1819mmas} plots the entropic variable $A$ versus the enclosed
mass fraction for the final collision products of cases 18, 19, and 20, as
determined both by SPH and {\sevensize MMAS}.  Cases 18 and 19 differ
only in the direction of approach of the third star, and
we again see that this variation has little effect on the SPH
results.  The kink in all of the profiles slightly inside $m/M=0.1$ marks the
boundary within which fluid from only the $0.8{\rm M}_\odot$ star
contributes.  {\sevensize MMAS} again reproduces this feature quite
well.  {\sevensize MMAS} does underestimate the shock heating to the core
and hence the central value of $A$, although some of this discrepancy is
due to the spurious heating evident in longer SPH simulations \citep{lom99}.
Nevertheless, it is likely that this difference between the
{\sevensize MMAS} and SPH models would last only as a transient during the
thermal relaxation in a stellar evolution calculation.  It is also worth
noting that, while
the SPH calculations need to be terminated before all of the
bound fluid can settle into equilibrium (see \S\ref{sph}),
the {\sevensize MMAS} $A$ profile does steadily increase outward throughout
the entire product.

\subsubsection{Sizes of collision products \label{sizes}}

Unfortunately it is a difficult task to determine the overall size of a
collision product, either with SPH simulations or with a package
like {\sevensize MMAS}.  Whenever there is any mass loss in
an SPH simulation, there will also be SPH particles that are nearly
unbound and, in practice, still moving away from the product when the
simulation is terminated.  These particles would ultimately form the
outermost layers of the collision product, but it would take an utterly
unfeasible amount of time to wait for them to come back and settle
into equilibrium.

The entropic variable $A$ profile produced by {\sevensize MMAS} seems quite
reasonable, both because it increases all the way out to the surface and
because the SPH results tend to approach its form as more
of the fluid settles into equilibrium.  However,
there are no simulation data to compare against for
the very outermost layers of a product and so the exact form of the profile there is
difficult to validate.  Not surprisingly, the radius of the collision
product is rather sensitive to the $A$ profile.  For example, simply
by changing the parameter $c_3$ from -1.0 to the
still very reasonable value of -1.1, which tends to distribute slightly more shock
heating to the outer layers \citep[see][]{lom02}, the radii of our
{\sevensize MMAS} final collision product models for case 1 and for
case 2 increase by about a factor of 2.  Despite such uncertainties,
it is still interesting to get a crude estimate of the sizes of the
collision products immediately from {\sevensize MMAS}.  In making
these estimates, we do not account for the expansion due to rotation,
but instead simply integrate the equation of hydrostatic equilibrium
for a non-rotating star with the same $A$ profile, using the outer
boundary condition that the pressure vanishes.  The radii calculated
therefore represent the sizes that the products would have if some
mechanism were to brake their rotation without disturbing their $A$
profiles.

\begin{figure*}
\includegraphics[width=100mm]{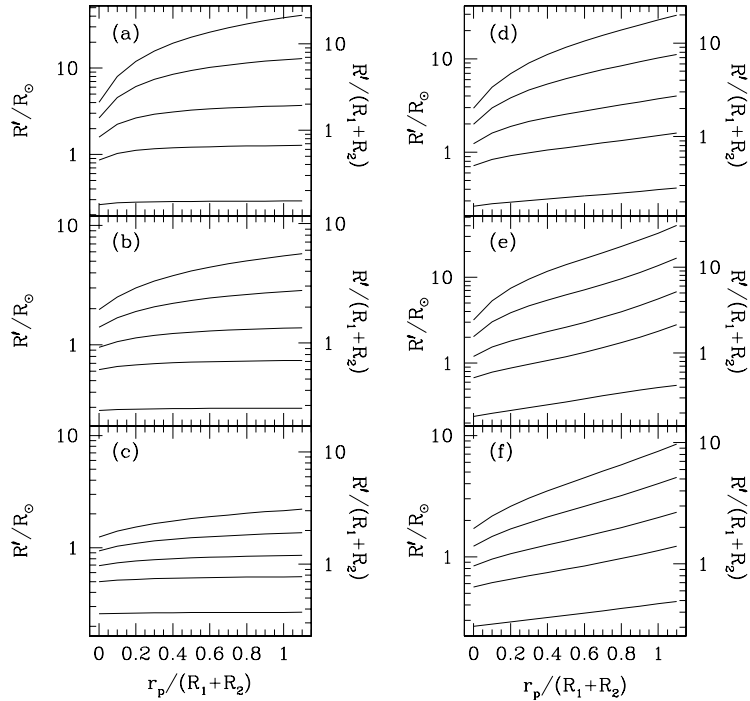}
\caption{As a function of the normalized periastron separation
$r_p/(R_1+R_2)$, each pane shows the radius $R^\prime$ that encloses, from
the top curve to the bottom one, 100, 99, 95, 86, and 50 per cent of
the total bound mass of the collision product, on a logarithmic scale and
as determined by {\sevensize MMAS}.  The scale on the left gives the radius
 in solar units, while the scale on the right normalizes the radius
to the sum of the parent star radii.
The combination of parent stars considered are (a)
0.8 and $0.8 {\rm M}_\odot$, (b) 0.6 and $0.6 {\rm M}_\odot$, (c) 0.4
and $0.4 {\rm M}_\odot$, (d) 0.8 and $0.6{\rm M}_\odot$, (e) 0.8 and
$0.4 {\rm M}_\odot$, and (f) 0.6 and $0.4 {\rm M}_\odot$.
\label{rvr}}
\end{figure*}

Fig.~\ref{rvr} plots the radii at various enclosed mass fractions for
%%%%%%%%%%%%%%%%%%%%%%%%%%%%%%%%%%%%%%%%%%%%%%%%%%%%%%%%%%%%%%%
% FIGURE 24
%%%%%%%%%%%%%%%%%%%%%%%%%%%%%%%%%%%%%%%%%%%%%%%%%%%%%%%%%%%%%%%
products generated in single-single star collisions involving 0.4, 0.6
and $0.8{\rm M}_\odot$ parent stars, as determined by {\sevensize
MMAS}.  These radii are plotted against the normalized periastron
separation $r_p/(R_1+R_2)$, which we allow to exceed unity slightly to
account for bulges in the parent stars.  The general trend is that as
the periastron separation increases, the collisions are more
long-lived, there is more shock heating, and the radii of the
collision products increase.  Because the fluid in the deep interior
of the product is largely shielded from shocks, the $A$ profile there,
and hence the radius $r$ profile, are not too strongly dependent on
the periastron separation of the collision.  As a result, the radii
versus periastron separation curves of Fig.~\ref{rvr} become closer
to horizontal as one looks to smaller enclosed mass fraction.   For
the cases examined in Fig.~\ref{rvr}, the full (100 per cent enclosed
mass) radius of the collision product is always at least about twice the sum
of the radii of the parent stars, and often even much larger than
this.  For example, if two $0.8{\rm M}_\odot$ stars suffer a grazing
($r_p\approx R_1+R_2$) collision, the collision product then has a
full radius of about $40{\rm R}_\odot$, about 20 times larger than the
sum $R_1+R_2$ of the parent star radii.  We therefore expect that the
collisional cross-section of these first products will be
significantly enhanced over that of their thermally relaxed
counterparts.

\begin{figure*}
\centering
\includegraphics[width=84mm]{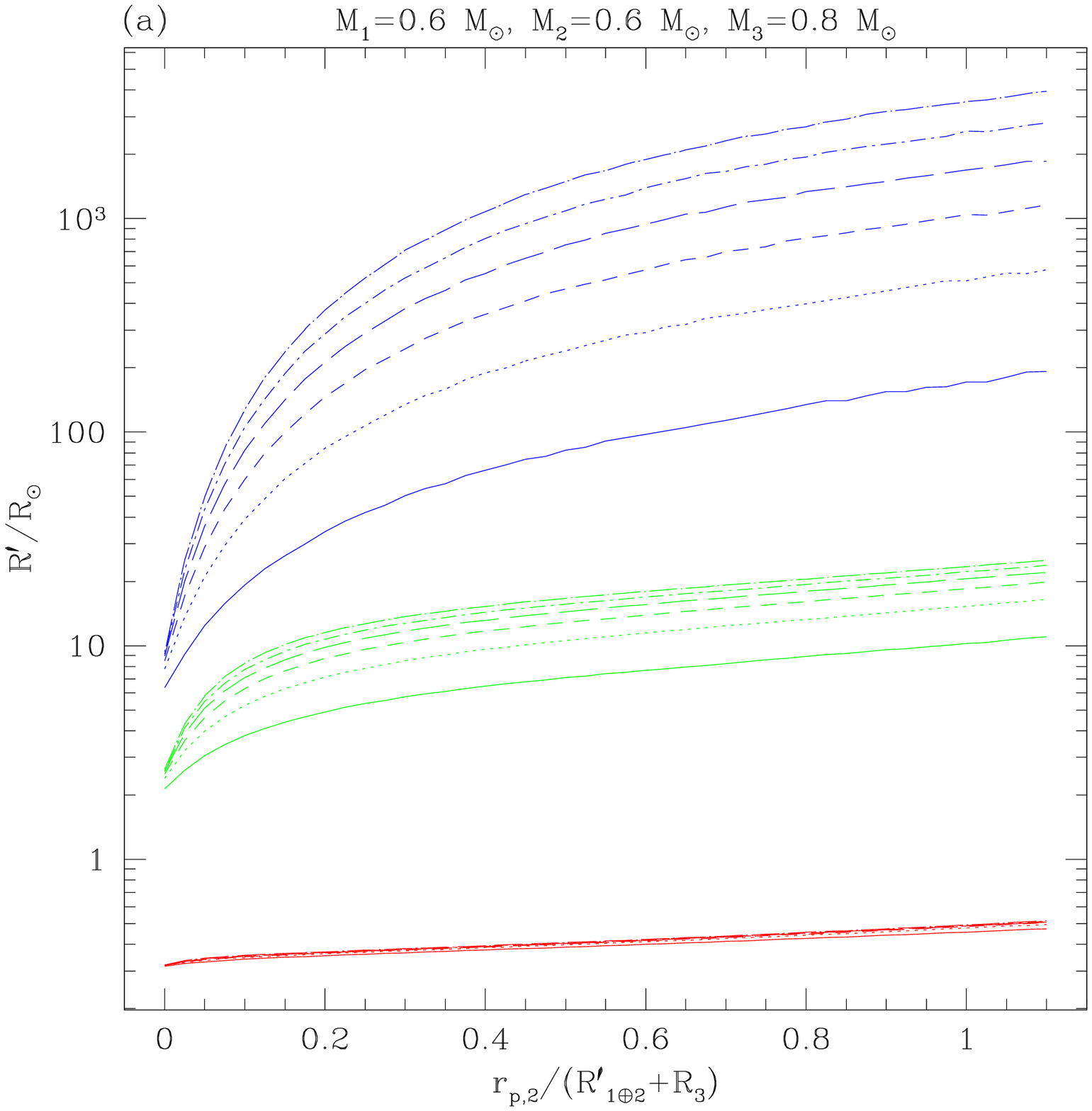}
\includegraphics[width=84mm]{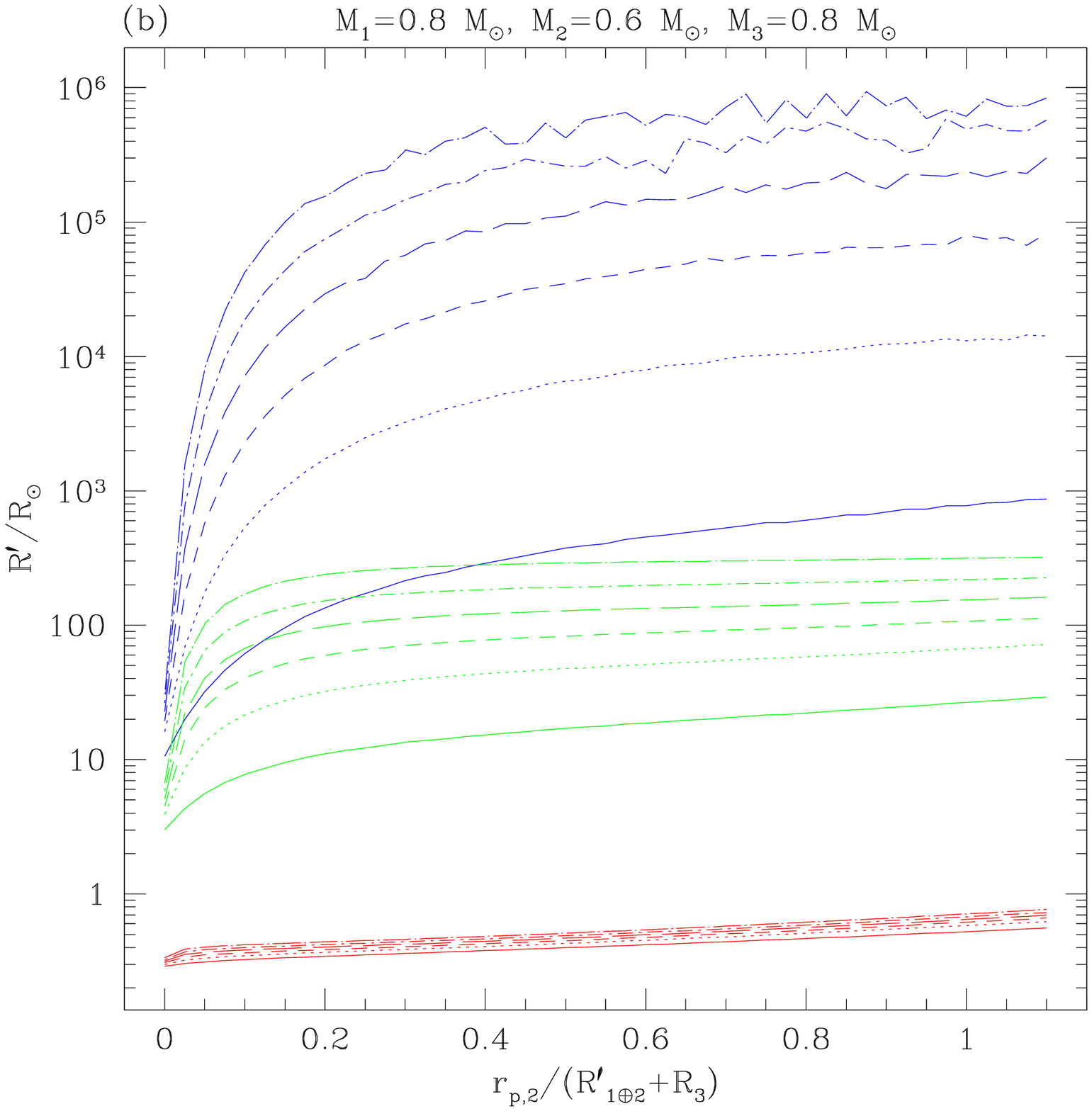}
\includegraphics[width=84mm]{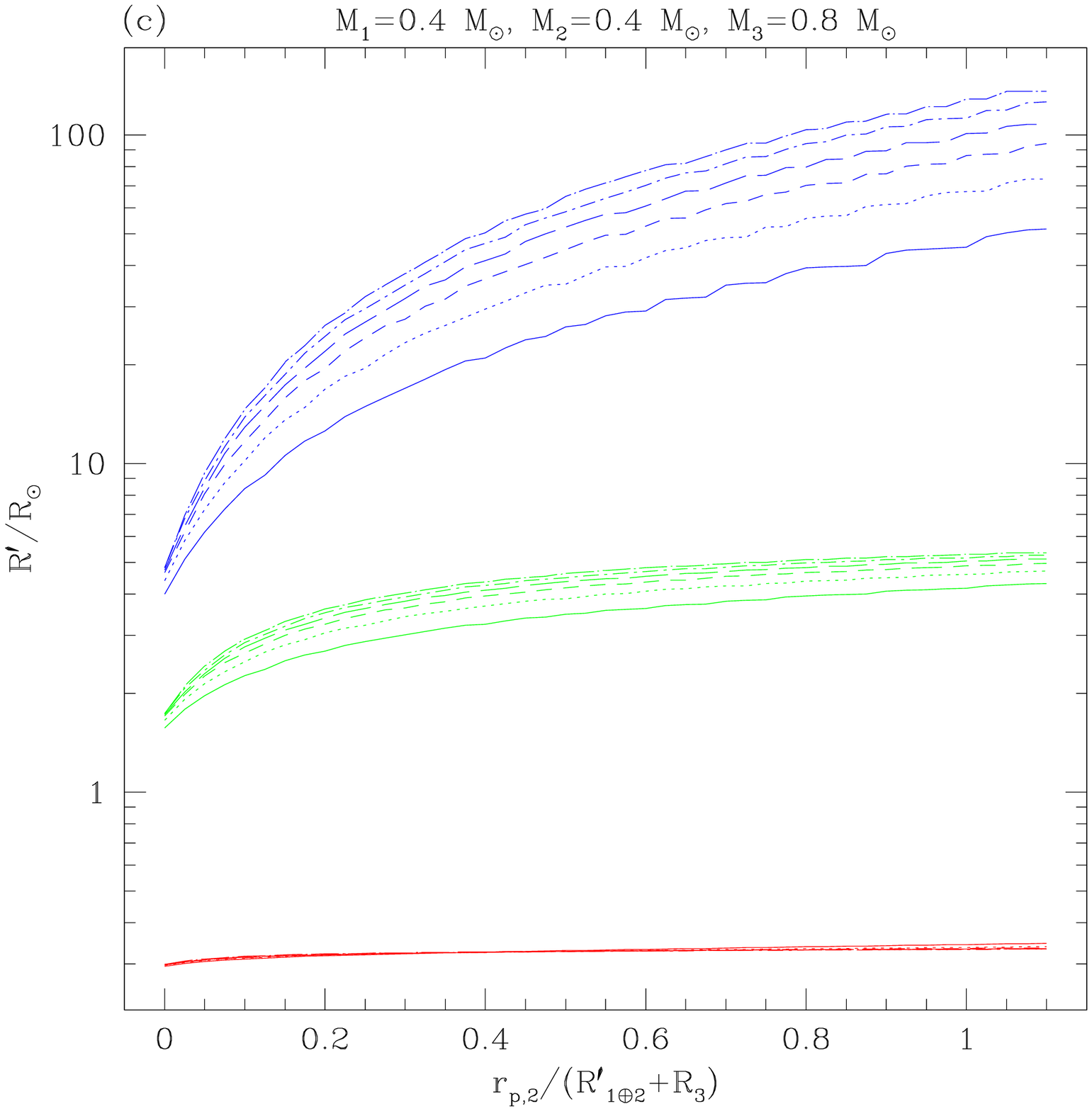}
\includegraphics[width=84mm]{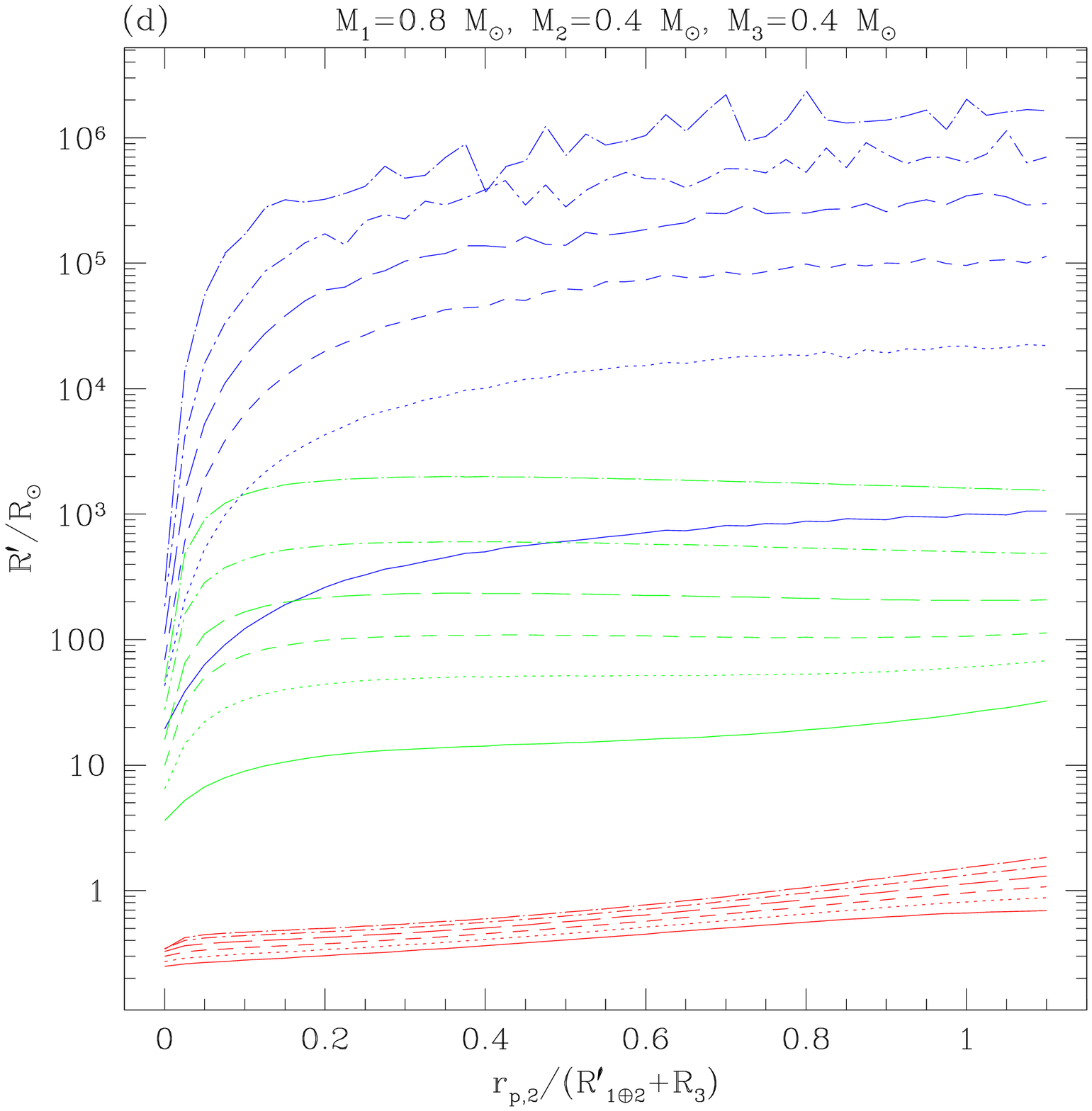}
\caption{
We plot
various radii $R^\prime$ of the final collision product (as determined
by {\sevensize MMAS}) as a function of the 
periastron separation of the second collision,
normalized to sum of the radii of the colliding stars.
The radius $R^\prime_{1\oplus 2}$ of the first collision product
is the 100 per cent radius from Fig.~\ref{rvr}
corresponding to various normalized periastron
separations for the first collision: $r_{p,1}/(R_1+R_2)=0$ (solid curve),
0.2 (dotted curve), 0.4 (short dashed curve), 0.6 (long dashed curve), 0.8 (dot - short
dashed curve), and 1.0 (dot - long dashed curve).  There are three curves of each
line type: the bottom one corresponds to the radius enclosing 50 per
cent of the final bound mass, the middle one corresponds to the 95 per cent
radius, and the top one to the 100 per cent radius.
\label{bw}} 
\end{figure*}

Fig.~\ref{bw} is similar to Fig.~\ref{rvr}, but for triple-star
%%%%%%%%%%%%%%%%%%%%%%%%%%%%%%%%%%%%%%%%%%%%%%%%%%%%%%%%%%%%%%%
% FIGURE 25
%%%%%%%%%%%%%%%%%%%%%%%%%%%%%%%%%%%%%%%%%%%%%%%%%%%%%%%%%%%%%%%
collisions.  We use different line types to represent various
normalized periastron
separations $r_p/(R_1+R_2)$ for the first collision, and
along the horizontal axis we vary the normalized periastron of the second
collision.  The curves give the radii at three
different enclosed mass fractions.  For each of the six $r_p$
values in a frame of
Fig.~\ref{bw}, we performed a nested loop over 45 equally spaced
values of the normalized periastron separation for the second collision,
from 0 to 1.1.  Therefore, {\sevensize MMAS} treated 270
different triple-star collisions (in a few minutes on
a Pentium IV workstation) for each of the four plots.  Note that there is a
general trend for the radius of the collision product to increase as
the first periastron separation increases, as expected; this
effect is mild for the 50 per cent enclosed mass radius, and
rather dramatic for the full radius.  Even more significant is the second periastron
separation, with grazing second collisions resulting in products that
are substantially larger than those from head-on collisions: the shock
heating suffered by the already diffuse outer layers of the first
collision product is severe when multiple pericentre passages occur
before merger.   Once $r_{p,2}$ grows large enough for the third
star's initial impact to be outside of the first product's core, so
that more than one pericentre passage would result before merger, then
the shock heating is no longer as sensitive to $r_{p,2}$ and the
full radius surfaces in Fig.~\ref{bw} tend to plateau.  How strongly
the full radius varies with the $r_{p,2}$ therefore
depends on the mass distribution within the first product.  For first
products with a more uniform density, such as in the product of two
$0.4{\rm M}_\odot$ stars, the final product size increases more
gradually and consistently with $r_{p,2}$.

As the mass of any one of the three parent stars is increased, the
trend is for the radius of the collision product to increase as well.
For example, Fig.~\ref{bw}(a) shows that for
collisions in which two $0.6{\rm M}_\odot$ stars collide and then a
$0.8{\rm M}_\odot$ collides with the first product, the final
collision product radius does not exceed a few times $10^3 {\rm
R}_\odot$.  If one of the $0.6{\rm M}_\odot$ stars is substituted with
a $0.8{\rm M}_\odot$ star, then the final radius can be as large as about
$10^6 {\rm R}_\odot$ [see Fig.~\ref{bw}(b)]. This extreme size is
due to the phenomenally diffuse outer layers of the product:
the average density of such a star is only $\sim 10^{-18}$ g cm$^{-3}$.
The noise visible on some of the full radius curves is
due to approaching the limiting numerical
precision during the structure integration in these diffuse regions.

From Fig.~\ref{bw} we see that the radius that encloses 95 per cent of
the total mass, while still large, is often orders of
magnitude smaller than the full radius of the final product.  Because of the low densities
involved, the full radius calculated is rather sensitive to the
details of the shock heating during the collision.  Changing the
{\sevensize MMAS} parameter $c_3$ from -1.0 to -1.1, for example, can increase
the full radius by a factor of a few, although the radius enclosing 50
per cent of the total mass does not change by more than a few per
cent.  Nevertheless, any reasonable form and amount of shock heating
yields products that are significantly larger than a thermally relaxed
star with the same mass and composition.

Colliding the same three parent stars in a different order does not
drastically affect the mass of the final product, although it does
significantly affect its size.  Consider, for example, 
frames (c) and (d) of Fig.~\ref{bw}.  If two $0.4{\rm M}_\odot$ stars collide and
then the resulting product collides with a $0.8{\rm M}_\odot$ star,
the final product typically has a radius of order $\sim 10 - 100 {\rm
R}_\odot$, but if the $0.8{\rm M}_\odot$ star is switched into the
first collision instead, the final radius is usually in the range
from  $\sim 100 - 10^6{\rm R}_\odot$.  The primary reason for this difference is that a
collision between the 0.4 and $0.8{\rm M}_\odot$ stars yields a
product with especially diffuse outer layers, and, as a result, is subject to a
larger number of passages and hence more shock heating during a second
collision.

\section{Discussion} \label{discussion}

\subsection{Concluding remarks}

We have used SPH and the software package {\sevensize MMAS} to study
triple-star collisions. Although such collisions span a tremendous
amount of parameter space, our modest number of SPH calculations do
provide some valuable insights.  For the (parabolic) encounters that
we consider, we find that the order in which stars collide (see
\S\ref{order}), the angle of approach of the third star
(\S\ref{direction}), and the periastron separation of the collisions
(\S\ref{spin}) have only a slight effect on the chemical composition
distribution within the final collision product.  The order and
orbital parameters of the collisions can, however, significantly affect
the size and structure of the product.

The results of \S\ref{comparison} help establish that the simple fluid
sorting algorithm of {\sevensize MMAS} reproduce the important
features of our SPH models, even when one of the parent stars is
itself a collision product.  The {\sevensize MMAS} package can
therefore be considered an adequate, if not an accurate, substitute
for a hydrodynamics code in many situations.  This realization will
help simplify the process of generating collision product models in
cluster simulations, because a full hydrodynamics calculation will not
necessarily need to be run for each collision.  Indeed, we hope the
{\sevensize MMAS} package will be used to help account for stellar
collisions in dynamics simulations of globular clusters.  Toward this
end, {\sevensize MMAS} is already being incorporated into two software
packages, {\sevensize TRIPTYCH} and {\sevensize TRIPLETYCH}, that
respectively treat encounters between two stars and among three stars
\citep[see][]{sil03}.\footnote{See {\tt
http://faculty.vassar.edu/lombardi/triptych/} and {\tt
http://faculty.vassar.edu/lombardi/tripletych/}.}  These packages are
controlled through a web interface and treat the orbital trajectories,
possible merger(s), and evolution of the merger product and therefore
incorporate three main branches of stellar astrophysics: dynamics,
hydrodynamics, and evolution.

The product size estimates of \S\ref{sizes} are admittedly crude.
For example, partial ionization and radiation pressure are neglected.
Although the exact size of a collision product is difficult to
determine, our calculations indicate that the first and final
collision products are always significantly larger than their
thermally relaxed counterparts would be.  Indeed, according to our
{\sevensize MMAS} calculations in \S\ref{sizes}, the final collision
product can have a radius up to $\sim 10^6 {\rm R}_\odot$, easily
exceeding the size of a typical red giant.
Furthermore, these calculations have assumed that some mechanism has
braked the often rapid rotation of the products, so any rotation that does
remain will only further enhance the size of the products.  The
extended sizes of the products will increase the multi-star collision
rate over that calculated in previous treatments of binary-single and
binary-binary encounters.

All of the scenarios we consider with SPH in this paper involve one
$0.8$ and 
two $0.6{\rm M}_\odot$ stars.  Without shock heating, the low-$A$ fluid of
the $0.8{\rm M}_\odot$ star would sink to the core of the final collision
product, while its high-$A$ portions of the $0.8{\rm M}_\odot$ would settle
in the outer layers.  The intermediate layers of the product would
consist of fluid with the same $A$ range from all the parents.  Simply
sorting the fluid in this way, without running a hydrodynamics
calculation, can therefore provide a zeroth-order model of the
collision product that captures some of its important  qualitative
features (see Fig.~\ref{zeroth_order}).

However, non-uniform shock heating during the collisions somewhat
alters the relative values of the entropic variable $A$ in the fluid,
resulting in a slightly different sorting pattern.  Because the amount
and distribution of shock heating are dependent on the details of a
collision, the sorting of the fluid varies with, for example, the
order in which the stars collide.  Shock heating can have larger
consequences on the chemical abundance profiles of elements, such as
C$^{13}$, that exist in substantial amounts only in a small shell in
the initial parent stars.  However, the chemical abundance profiles of
most elements, particularly helium, are always qualitatively the same,
regardless of how the three stars are merged.  Because the abundance
and distribution of helium (and hence hydrogen) is one of the most
important factors in determining the collision product's subsequent
course of stellar evolution, we believe that the order and geometry of
the collisions will not significantly affect the stellar evolution of
the product.  Indeed, \citet{sil02} have recently presented a set of
stellar evolution calculations for a collision product for which the
starting {\sevensize YREC} models were generated from SPH calculations
of different resolutions.  The variations in the initial helium
profiles of their models are roughly comparable to those in our helium
profiles resulting from colliding three parent stars in different
ways.  Although \citet{sil02} do find detailed differences in the
evolution, especially during the ``pre-main-sequence'' contraction,
the evolutionary tracks and time-scales are quite similar.  We
therefore feel that, for low-velocity collisions, the hydrodynamical
details of how three stars are merged will not significantly affect
the stellar evolution of the collision product---the major caveat here
being that the geometry of the collisions can of course affect the
rotation of the product, which in turn can greatly affect its
evolution \citep{sil01}.

Surface abundances of lithium and beryllium are particularly
interesting to monitor, as these elements can be used as observational
indicators of mixing and perhaps collisional history.  As in the single-single star collisions
presented by \cite{lom02}, we find that the triple-star mergers
presented here yield collision products that are severely depleted of
lithium and somewhat of beryllium at the surface.  Even in the
relatively gentle (parabolic) cases that we have considered, the
collisions are energetic enough to expel most of the lithium and
beryllium from the outer layers of the parents.

\subsection{Future work \label{future}}

There are many scenarios to explore when dealing with collisions in
environments as chaotic as dense stellar systems.
Different orbital geometries besides the parabolic trajectories
treated in this paper still need to be considered in more detail.  Large stellar velocities in
galactic nuclei lead to hyperbolic collisions.  In globular
clusters, perturbations to a binary can lead to an elliptical collision,
while an encounter with a very hard binary can lead to significantly
hyperbolic collisions.

Future studies may want to include a more detailed look at the
hydrodynamics during grazing encounters, which could be done
efficiently with the help of GRAPE (short for GRAvity PipE) special
purpose hardware for calculating the self-gravity of the system.
Furthermore, encounters involving more than three stars, such as in
binary-binary interactions, may also warrant further examination: for
example, the final collision product generated in a triple-star merger
is typically so extended that it could immediately start suffering
Roche lobe overflow if left in orbit around a fourth star.

Collisions among a larger variety of stellar types and masses,
reflective of the diverse
populations of clusters, will also need to be explored.  We have been
concentrating on low-mass main-sequence stars, but collisions between
high-mass main-sequence stars in young compact star clusters, or
giants located in the dense cores of globular clusters, for example,
are frequent.  A logical first step would be to examine high-mass
main-sequence stars in a runaway merger scenario.  It would therefore
be very useful to develop a generalization of the fluid sorting
method that includes radiation pressure in the equation of state.

Due to shock heating during the collision, the product is much larger
than a thermally equilibrated main-sequence star of the same mass.
How much of an effect this increased radius has on the effective cross
section for merger is subject to many variables, including the
structure of the product's outer layers and the velocity of
approaching stars.  In environments such as active galactic nuclei,
where relative velocities tend to be high, the low-density outer
layers of a newly formed collision product could likely get stripped
by passing stars.  However, in globular clusters, where stellar
velocities tend to be small, collisions with even low-density
envelopes may lead to significantly increased rates of merger.  It
would be useful to develop a robust collision module that could
quickly predict whether any given collision trajectory will lead to a
merger, and, if not, describe how the stars are affected by the
interaction.

One simple approximation often implemented in cluster simulations is
that a collision product instantaneously achieves its thermally
relaxed radius, a good approximation when the time between collisions
is much longer than the thermal time-scale.   Arguing instead that the
global thermal time-scale of the first product can be much larger than
the time between collisions in interactions involving binaries, we
make a different approximation in this paper, namely that the first
product's radius (and more generally its structure) does not
substantially evolve between collisions.
Future scattering experiments could
model thermally relaxing stars and study more carefully the timescale
between collisions mediated by binaries.  The thermal
time-scale in the outer layers of a collision product can be orders of
magnitude less than its global thermal time-scale \citep[see table 1
of][]{1997ApJ...487..290S}, so that it may actually be necessary to
follow the thermal contraction and stellar orbits simultaneously.
Indeed, in the extremely low density layers of a collision product,
it is even possible for the thermal time-scale to be comparable to the
(hydro)dynamical time-scale, so that the product could undergo
significant thermal contraction even before it reaches hydrodynamical
equilibrium.  It would be helpful if future stellar evolution
calculations of collision products included a detailed description
of the products' size and structure throughout the thermal relaxation
stage.  How quickly the outer layers of the thermally expanded product
change with time will substantially affect its likelihood of
subsequent collisions.  Initial conditions for such stellar evolution
calculations could be provided by the publicly available {\sevensize
MMAS} package.

The primary hurdle for incorporating collisions into realistic
stellar dynamics simulations is currently the stellar evolution of the
collision products.  Such stars are highly non-canonical,
typically with very peculiar structural and composition profiles,
and present a challenging set of initial conditions for stellar
evolution codes.  To make matters even more intricate, rotation, which
is typically rapid
after merging, will affect the structural properties and chemical
compositions
of the stars as they evolve \citep[e.g.,][]{sil01}.  This rapid
rotation also has the effect of ejecting mass as the product thermally
contracts.  Studying this emitted mass will be
worthwhile, as it may likely carry away angular momentum and at least
partially brake rotating collision products.

\section*{Acknowledgments}

We would like to thank Fred Rasio for helpful comments and the use of
his SPH code, Josh Faber for having parallelized this code, Randall
Perrine for assistance in preliminary SPH calculations, Alison
Sills for providing {\sevensize YREC} models for the parent stars, and
the referee Marc Freitag for valuable comments that helped
improve this paper.  We are also grateful to the participants of the first
two MODEST workshops for useful discussions, especially 
Jarrod Hurley, Piet Hut, Steve McMillan, Onno Pols, Simon Portegies
Zwart, and Peter Teuben.  This work was supported by NSF Grants
AST-0071165, MRI-0079466, and AST-0205991.  This work was also
supported by the National Computational Science Alliance
under Grant AST980014N and utilized the NCSA SGI/Cray Origin2000
parallel supercomputer.

%\bsp

\label{lastpage}

\end{document}